\documentclass[preprint,aps,preprintnumbers,amsmath,amssymb]{revtex4}

\usepackage{sidecap}
\usepackage{color}
\usepackage[]{graphicx}
\usepackage{latexsym}
\usepackage{natbib}
\usepackage{amsmath}
\usepackage[caption=false]{subfig}
\usepackage{multirow}
\usepackage{mathtools}
\usepackage{algorithm, algpseudocode}
\usepackage{textcomp}
\usepackage{rotating}
\usepackage{tikz}
 
\usepackage{array}
\usepackage{longtable}
\newcolumntype{H}{>{\setbox0=\hbox\bgroup}c<{\egroup}@{}}

\PassOptionsToPackage{hyphens}{url}
\usepackage[hyphens]{url}
\usepackage[breaklinks=true]{hyperref}

\newcommand{\down}{\downarrow}
\usepackage[final]{feynmp}
\begin{document}


\title{Fibration symmetries uncover the building blocks of biological
  networks}

\vspace{-.5cm}

\author{Flaviano Morone, Ian Leifer, Hern\'an A. Makse}

\vspace{-.5cm}

\affiliation{Levich Institute and Physics Department, City
  College of New York, New York, NY 10031}

\vspace{-.5cm}

\begin{abstract}




  





A major ambition of systems science is to uncover the building blocks
of any biological network to decipher how cellular function emerges
from their interactions. Here, we introduce a graph representation of
the information flow in these networks as a set of input trees, one
for each node, which contains all pathways along which information can
be transmitted in the network. In this representation, we find
remarkable symmetries in the input trees that deconstruct the network
into functional building blocks called fibers. Nodes in a fiber have
isomorphic input trees and thus process equivalent dynamics and
synchronize their activity. Each fiber can then be collapsed into a
single representative base node through an information-preserving
transformation called 'symmetry fibration', introduced by Grothendieck
in the context of algebraic geometry. We exemplify the symmetry
fibrations in gene regulatory networks and then show that they
universally apply across species and domains from biology to social
and infrastructure networks. The building blocks are classified into
topological classes of input trees characterized by integer branching
ratios and fractal golden ratios of Fibonacci sequences representing
cycles of information. Thus, symmetry fibrations describe how
complex networks are built from the bottom up to process information
through the synchronization of their constitutive building blocks.

\end{abstract}

\maketitle

A central theme in systems science is to break down the system into
its fundamental building blocks to then uncover the principles by
which complex collective behavior emerges from their interactions
\cite{leibler,alon,gellmann}. In number theory, every natural number
can be represented by a unique product of primes. Thus, prime numbers
are the building blocks of natural numbers. This mathematical notion
of building blocks is extended to the more abstract notion of group
theory since
finite groups can also be factored into simple subgroups~\cite{dixon}.
The latter example, entirely abstract as it may be, has important
implications for natural systems due to the fundamental relationship
between group theory and the notion of symmetry, that has led to the
discovery of the fundamental building blocks of matter, such as quarks
and leptons ~\cite{gellmann,weinberg}. Here we ask whether similar
principles of symmetry can uncover the fundamental building blocks of
biological
networks~\cite{leibler,alon,alon-science,caldarelli}. Primary examples
of these networks are gene regulatory networks that control gene
expression in cells~\cite{alon,alon-ecoli,karlebach,klipp}, as well as
metabolic networks, cellular processes and pathways, neural networks
and ecosystems and, beyond biology, to other information-processing
networks like social and infrastructure networks
\cite{caldarelli}. Previous studies have identified building blocks or
`network motifs' ~\cite{alon,alon-science,alon-ecoli} by looking for
patterns in the network that appear more often that they would by pure
chance. The crux of the matter is to test whether the building blocks
of these networks obey a predictive design principle that explains how
the cell functions, and whether such a principle can be expressed in
the language of symmetries.



We introduce the use of symmetries in biological networks by analyzing
the transcriptional regulatory network of bacterium {\it Escherichia
  coli} ~\cite{regulon}, since this is a well-characterized
network. We find that this network exhibits {\bf fibration symmetries}
~\cite{grothendieck,boldi,golubitsky}; first introduced by
Grothendieck~\cite{grothendieck} in the context of algebraic geometry.



Symmetry fibrations are morphisms between networks that identify
clusters of synchronized genes (called {\bf fibers}) with isomorphic
input trees.
Genes in a fiber are collapsed by a symmetry fibration into a single
representative gene called the {\bf base}. The fibers are then the
synchronized building blocks of the genetic network and symmetry
fibrations are transformations that preserve the dynamics of
information flow in the network. We use this symmetry principle to
classify the building blocks into topological classes of input trees
characterized by integer branching ratios and complex topologies with
golden ratios of Fibonacci sequences representing cycles in the
network. We then show that symmetry fibrations explain synchronization
patterns of gene co-expression in cells and universally apply to a
range of complex networks across different species and domains beyond
biology.

\section{Results}

We search for symmetries in the {\it E. coli} transcriptional
regulatory network (most updated compilation at RegulonDB
~\cite{regulon}) where nodes are genes and a directed link represents
a transcriptional regulation (see Supplementary Information
Section~\ref{transcriptional}).

A directed link from a source gene $i$ to a target gene $j$ in a
transcriptional regulatory network represents a direct interaction
where gene $i$ encodes for a transcription factor that binds to the
binding site of gene $j$ to regulate (activate or repress) its
expression. Such a link represents a regulatory 'message' sent by the
source to the target gene using the transcription factor as a
`messenger'.
This process defines the 'information flow' in the system which is not
restricted to two interacting genes, but it is transferred to
different regions within the network that are accessible through the
connecting pathways.
The information arriving to a gene contains the entire history
transmitted through all pathways that reach this gene. We formalize
this process of communication between genes with the notion of
{\it 'input tree'} of the gene.
In a network $G=(N_G, E_G)$ with $N_G$ nodes and $E_G$ directed edges,
for every gene $i\in N_G$ there is a corresponding input tree, denoted
as $T_i$, which is the tree of all pathways of $G$ ending at $i$. More
precisely, $T_i$ is a rooted tree with a selected node $i$ at the
root, such that every other node $j$ in the tree represents the
initial node of a path in the network ending at $i$.

Next, we analyze the input trees in the {\it E. coli} sub-circuit
shown in Fig. \ref{cpxr}a regulated by gene {\it cpxR} which regulates
its own expression (via an autoregulation activator loop) and also
regulates other genes as shown in the figure.  Gene {\it cpxR} is not
regulated by any other transcription factor in the network, so,
we say that this gene forms its own `strongly connected component',
see below. Therefore, it is an ideal simple circuit to explain the
concept of fibration.

\subsection{Input tree representation}

In practice, the input tree of a gene is constructed as follows (SI
Section~\ref{sec:input}). Consider the circuit in
Fig. \ref{cpxr}a. The input tree of gene {\it spy} depicted in
Fig. \ref{cpxr}b starts with {\it spy} at the root (first
layer). Since this gene is upregulated by {\it baeR} and {\it cpxR},
then, the second layer of the input tree contains these two pathways
of length one starting at both genes. Gene {\it baeR} is further
upregulated by {\it cpxR} and by itself through the autoregulation
loop and {\it cpxR} is also autoregulated. Thus, the input tree
continues to the third layer taking into account these three possible
pathways of length 2, one starting at {\it baeR} and two starting at
{\it cpxR}. The procedure now continues, and since there are loops in
the circuit, the input tree has an infinite number of layers.

The input tree formalism is a powerful framework to search for
symmetries in information-processing networks, in that it replaces the
canonical notion of a single trajectory with the set of all possible
`histories' from an initial to a final state of the network, and this
makes, in practice, reasonably straightforward to `guess' a type of
symmetry which is not apparent in the classical network framework.
Based on results from \cite{boldi,golubitsky,deville,sanders}, we will
show in Section \ref{sync} that if two input trees have the same
'shape', then the genes at the root of the input trees synchronize
their activity
\cite{other0,pecora1,pecora2,stewart,arenas,kurths,strogatz}, even
though their input trees are made of different genes.
This informal notion of equivalence is formalized by isomorphisms. An
isomorphism between two input trees is a bijective map that preserves
the topology of the input trees including the type of
links. Specifically, a map $\tau: T\to T'$ is an isomorphism iff for
any pair of nodes $a$ and $b$ of $T$ connected by a link, the pair of
nodes $\tau(a)$ and $\tau(b)$ of $T'$ is connected by the same type of
link (SI Section~\ref{sec:isomorphic}). In practice, this means that
isomorphic input trees are `the same' except for the labeling of the
nodes. Genes with isomorphic input trees are symmetric and
synchronous. We quantify this result, next, by introducing the concept
of symmetry fibration \cite{boldi}.






\subsection{Symmetry fibration of a network}

The set of all input tree isomorphisms defines the symmetries of the
network, which can be described by a {\it 'Grothendieck
  fibration'}~\cite{grothendieck}.  The original Grothendieck
definition of fibration is between categories~\cite{grothendieck}, so
the passage to a definition of fibrations between graphs requires to
associate a category with a graph and rephrase Grothendieck's
definition in elementary terms. Different categories may be associated
with a graph, giving rise to different notions of fibrations between
graphs. The notion of fibration that we use henceforth has been
introduced in computer science as a {\it 'surjective minimal graph
  fibration'} ~\cite{boldi,deville}.

In general, a graph fibration $G = (N_G, E_G)$ is any morphism
\begin{equation}
\psi: G\to B
\end{equation}
that maps $G$ to a graph $B = (N_B, E_B)$ (with $N_B$ nodes and $E_B$
edges) called the {\it 'base'} of the graph fibration $\psi$ (SI
Section~\ref{sec:base}).
In this work we consider a surjective minimal graph fibration
\cite{boldi} which is a graph fibration $\psi$ that maps all nodes
with isomorphic input trees inside a fiber to a single node in $B$,
thus producing the minimal base of the network.
In this case, the base $B$ consists of a graph where all genes in a
fiber have been collapsed into one representative node by the minimal
fibration.  Thus, a surjective minimal graph fibration, hereafter
called symmetry fibration for the sake of lexical convenience, leads
to a dimensional reduction of the network into its irreducible
components.  Crucially, a symmetry fibration is a dimensional
reduction that preserves the dynamics in the network as we show next.

\subsection{Symmetry fibration leads to synchronization}
\label{sync}

Next, we explain the connection between fibration and synchrony in a
generality that is needed to justify our results following
Ref. \cite{deville,sanders}.  In order to describe the dynamical state
of each gene in the transcriptional regulatory network, we first
attach a phase space to each node in $G=(N_G, E_G)$ by considering a
map $P:N_G \rightarrow M$ that assigns each node $i \in N_G$ to the
phase space of the node denoted by the manifold $M$. For example, in a
transcriptional regulatory network we assign to each gene $i \in N_G$
the phase space of real numbers $M = \mathbb{R}$. Then, the state of
each gene is described by $x_i(t) \in \mathbb{R}$, representing the
expression level of the gene $i$ at time $t$, which is typically
measured by mRNA concentration of gene product.
We then obtain the total phase space of $G$ as the product $PG =
\prod_{i \in N_G}P(i)$.

The fibers partition the graph $G$ into unique and non-overlapping sets
$\Pi = \{\Pi_1, \dots, \Pi_r\}$, such that $\Pi_1 \cup \dots \cup
\Pi_r = G$ and $\Pi_k \cap \Pi_l = \emptyset$ if $k \neq l$
\cite{cardon}.  We denote $i \sim_\Pi j$ when the input-trees of $i$
and $j$ are isomorphic and belong to the same fiber $\Pi_k$ . That is,
$\exists k \, \mid \, i, j \in \Pi_k$ and there exist a symmetry
fibration that sends both nodes to the same node in the base, $\psi(i)
= \psi(j)$.
DeVille \& Lerman \cite{deville} showed that symmetry fibrations
induce robust synchronization in the system (Theorem 4.3.1 in
\cite{deville}). In particular, it was shown that if $\psi$ is a
symmetry fibration then--- by proposition 2.1.12 in
Ref. \cite{deville}--- there exist a map
$\mathbb{P}_\psi:PB\rightarrow PG$ that maps
the total phase space of the base $B$, named $PB$, to the total phase
space of the graph $G$.  This map 
creates a polysynchronous subspace of synchronized solutions in fibers:
$\Delta_\Pi=\{x \in PG \, \mid \, x_{i}(t) = x_{j}(t) \,\, \mbox{\rm
  whenever} \,\, \psi(i) = \psi(j)\}$, where each set of synchronous
components of this subspace corresponds to a fiber in $\Pi$ (Lemma
5.1.1 in \cite{deville}, see also \cite{sanders}). In other words,
$\Delta_\Pi$ is a polysynchronous subspace of $PG$, such that
components $x_i, x_j \in x$ synchronize (i.e., $x_i(t) = x_j(t)$)
whenever the symmetry fibration $\psi$ sends them to the same node in
$B$.

According to these results, we interpret synchronous genes to process
the same information received through isomorphic pathways in the
network, and, accordingly, we interpret a symmetry fibration as a
transformation that preserves the dynamics of information flow since
it collapses synchronous nodes in fibers (redundant from the point of
view of dynamics) into a common base with identical dynamics as the
fiber.

Synchronous nodes in a fiber induced by symmetry fibrations correspond
to the 'minimal balanced coloring' in ~\cite{golubitsky}. A balanced
coloring assigns two nodes the same color only if their inputs,
self-consistently, receives the same content of colored nodes, whence
the term `balanced'. Thus, the flow of information arriving to genes
in a fiber is analogous to a process of assigning a color to each gene
such that each gene `receives' the colors from adjacent genes via
incoming links and `sends' its color to the adjacent genes via its
outgoing links.  The nodes in a fiber have the same color symbolizing
the fact that they synchronize.  The nodes with the same color in the
balanced coloring partition ~\cite{golubitsky} correspond to fibers
induced by symmetry fibrations \cite{deville}. We use the minimal
balanced coloring algorithm proposed in \cite{kamei} for the
computation of minimal bases \cite{cardon} to find fibers (SI Section
~\ref{algorithm1}).

\subsection{Strongly connected components of the {\it E. coli} network}


The input trees in the {\it E. coli} {\it cpxR} circuit are displayed
in Fig.~\ref{cpxr}b. The input trees of {\it baeR} and {\it spy} are
isomorphic
and define the {\it baeR}-{\it spy} fiber (Fig.~\ref{cpxr}c). We call
this circuit a feed-forward fiber (FFF).  The input tree of {\it cpxR}
is not isomorphic to either {\it baeR} or {\it spy}, and therefore
{\it cpxR} is not symmetric with these genes, but it is isomorphic to
{\it bacA, slt} and {\it yebE} forming another fiber.  Likewise, genes
{\it ung, tsr} and {\it psd} are all isomorphic composing another
fiber (Fig. \ref{cpxr}b). Figure \ref{cpxr}d shows the symmetry
fibration $\psi: G\to B$ that collapses the genes in the fibers to the
base $B$. Figure \ref{cpxr}e shows another example (out of many) of a
single connected component, {\it fadR}, and its corresponding
isomorphic input trees (Fig. \ref{cpxr}f), fibers and base.

The dynamical state of a gene is encoded in the topology of the
input-tree. In turn, this topology is encoded by a sequence, $a_i$,
defined as the number of genes in each $i-$th layer of the input tree
(Fig. \ref{cpxr}b). The sequence $a_i$ represents the number of paths
of length $i-1$ that reach the gene at the root.  This sequence is
characterized by the branching ratio $n$ of the input tree defined as
$a_{i+1}/a_i \xrightarrow[i \to \infty]{} n$, which represents the
multiplicative growth of the number of paths across the network
reaching the gene at the root.  For instance, the input trees of genes
{\it baeR-spy} (Fig. \ref{cpxr}b) encode a sequence $a_i=i$ with
branching ratio $n=1$ representing the single ($n$=1) autoregulation
loop inside the fiber.




Beyond several single-gene strongly connected components like those
shown in Fig. \ref{cpxr}, we find that the {\it E. coli} network has
other strongly connected components [in a strongly connected
  component, each gene is reachable from every other gene, SI Section
  \ref{gcc}], three in total, which regulate more involved topologies
of fibers. We find: (i) a two-gene strongly connected component
composed of master regulators {\it crp-fis} involved in a myriad of
functions like carbon utilization (Fig. \ref{components}a, top), (ii)
a five-gene strongly connected component involved in the stress
response system (SI Fig. \ref{soxr}), and (iii) the largest strongly
connected component at the core of the network which is composed of
genes involved in the pH-system that regulate hydrogen concentration
(Fig. \ref{components}b). Each of these three components regulate a
rich variety of fiber topologies which are collapsed into the base by
the symmetry fibration $\psi: G \to B$, as shown in the figure.

\subsection{Fiber building blocks}

We find that the transcriptional regulatory network of {\it E. coli}
is organized in 91 different fibers.  The complete list of fibers in
{\it E. coli} is shown in SI Section \ref{stats} and SI-Table
\ref{all_building_blocks} and the statistics are shown in SI Table
\ref{statistics}. Plots of each fiber are shown in Supplementary File
1.  We find a rich variety of topologies of the input trees. Despite
this diversity, the input trees present common topological features
that allow us to classify all fibers into concise classes of
fundamental 'fiber building blocks' (Figs. \ref{blocks}a and
\ref{blocks}b).
We associate a building block to a fiber by considering the genes in
the fiber plus the external in-coming regulators of the fiber plus the
minimal number of their regulators in turn that are needed to
establish the isomorphism in the fiber. When the fiber is connected to
any external regulator, either via a direct link or through a path in
the strongly connected component forming a cycle, then the genes in
this cycle are considered part of the building block of the fiber,
since such a cycle is crucial to establish the dynamical
syncronization state (when there is more than one cycle, the shortest
cycle is considered).


We find that the most basic input tree topologies can be classified by
integer 'fiber numbers' $| n, \ell\rangle$ reflecting two features:
(a) infinite $n$-ary trees with branching ratio $n$ representing the
infinite pathways going through $n$ loops inside the base of the
fiber, and (b) finite trees representing finite pathways starting at
$\ell$ external regulators of the fiber.
The most basic fibers in {\it E. coli} have three values of $n=0, 1,
2$ (Fig. \ref{blocks}a): {\it (i)} fibers with $n=0$ loops, called
Star Fibers (SF), {\it (ii)} fibers with $n=1$ loop, called Chain
Fibers (CF), and {\it (iii)} fibers with $n=2$ loops, called
Binary-Tree Fibers (BTF). This classification does not take into
account the types of repressor or activator links in the building
blocks, which lead to further sub-classes of fibers that determine the
type of synchronization (fixed point, limit cycles, etc) and thus the
functionality of the fibers.





 
Figure \ref{blocks}a shows a sample of dissimilar circuits that can be
concisely classified by $ |n, \ell\rangle$ (full list in Supplementary
File 1). For instance the $n=0$ SF class includes dissimilar circuits
like $|\mbox{\it arcZ-ydeA}\rangle = |0, 1\rangle$,
$|\mbox{\it dcuC-ackA}\rangle = |0, 2\rangle$ which is a bi-fan
network motif \cite{alon}, and generalizations with $\ell=3$
regulators like $|\mbox{\it dcuR-aspA}\rangle = |0, 3\rangle$
(Fig. \ref{blocks}a, top).
The main feature of these building blocks is that they do not contain
loops and therefore the input trees are finite. The CF class contains
$n=1$ loop in the fiber, and therefore an infinite chain in the input
tree, like the autoregulated loop in the fiber $|\mbox{\it
  ttdR}\rangle = |1, 0 \rangle$. We note that while the
  input tree is infinite, the topological class is characterized by a
  single number $n=1$ concisely represented in the base. Furthermore,
  a theorem proven by Norris ~\cite{norris} demonstrates that it
  suffices to test $N_G-1$ layers of the input trees to prove
  isomorphism, even though the input tree may contain an infinite
  number of layers. Adding one
external regulator ($\ell = 1$) to this circuit, converts it to
the purine fiber $|\mbox{\it purR}\rangle = |1, 1\rangle$ which is an
example of a FFF, like the {\it baeR} circuit in Fig. \ref{cpxr}a.
This circuit resembles a feed-forward loop motif \cite{alon}, but it
differs in the crucial addition of the autoregulator loop at {\it
  purR} that allows genes {\it purR} and {\it pyrC} to synchronize.
When another external regulator is added, we find the idonate fiber
$|\mbox{\it idnR}\rangle = |1, 2\rangle$.
More elaborated circuits contain two autoregulated loops and feed-back
loops featuring trees with branching ratio $n=2$.

\subsection{Fibonacci fibers}
\label{ff}

So far we have analyzed building blocks that receive information from
the external regulators in their respective strongly connected
components, but do not send back information to the external
regulators. These topologies are characterized by integer branching
ratios, $n=0, 1, 2$, as shown in Fig. \ref{blocks}a. We find, however,
more interesting building blocks that also send information back to
their regulators.
These circuits contain additional cycles in the building blocks that
transform the input trees into fractal trees characterized by
non-integer fractal branching ratios.
Notably, the building block of the fiber {\it uxuR-lgoR} that is
regulated by the connected component {\it crp-fis} (Fig.
\ref{components}) forms an intricate input tree (Fig. \ref{blocks}b,
top) where the number of paths of length $i-1$ is encoded in a
Fibonacci sequence: $a_i = $1, 3, 4, 7, 11, 18, 29, ...  characterized
by the Fibonacci recurring relation: $a_1=1$, $a_2=3$, and
$a_i=a_{i-1} + a_{i-2}$ for $i>2$. This sequence leads to the
non-integer branching ratio known as the golden ratio: $a_{i+1}/a_i
\xrightarrow[i \to \infty]{} \varphi = (1+\sqrt{5})/2 = 1.6180...$

This topology arises in the genetic network due to the combination of
two cycles of information flow. First, the autoregulation loop inside
the fiber at {\it uxuR} creates a cycle of length $d=1$ which
contributes to the input tree with an infinite chain with branching
ratio $n=1$. This sequence is reflected in the Fibonacci series by the
term $a_i = a_{i-1}$.  The important addition to the building block is
a second cycle of length $d=2$ between {\it uxuR} in the fiber and its
regulator {\it exuR}: {\it uxuR $\to$ exuR $\to$ uxuR}. This cycle
sends information from the fiber to the regulator and back to the
fiber by traversing a path of length $d=2$ that creates a 'delay' of
$d=2$ steps in the information that arrives back to the fiber (see
Fig. \ref{blocks}b, top). This short-term 'memory' effect is captured
by the second term $a_i = a_{i-2}$ in the Fibonacci sequence leading
to $a_i = a_{i-1} + a_{i-2}$ and the golden ratio. We call this
topology a Fibonacci fiber (FF).

This argument implies that an autoregulated fiber that further
regulates itself by connecting to its connected component via a cycle
of length $d$ encodes a generalized Fibonacci sequence of order $d$
defined as $a_i = a_{i-1} + a_{i-d}$ with generalized golden ratio
$\varphi_d$ (Fig. \ref{blocks}b fourth row). We find such a Fibonacci
sequence in the {\it evgA-nhaR} fiber building block linked to the pH
strongly connected components shown in Fig. \ref{components}b. This
fiber contains an autoregulation cycle inside the fiber and also an
external cycle of length $d=4$ through the pH strongly connected
component: {\it evgA $\to$ gad E $\to$ gadX $\to$ hns $\to$ evgA}
(Fig. \ref{blocks}b, third row).  This topology forms a fractal input
tree with sequence $a_i = a_{i-1} + a_{i-4}$ (sequence A123456 in
\cite{sloane}) and branching golden ratio $\varphi_4=1.38028...$ We
call this topology 4-Fibonacci fiber, 4-FF.  Generalized Fibonaccis
appear inside strongly connected components, like the {\it rcsB-adiY}
3-FF in the pH system (Fig. \ref{blocks}b, second row). Likewise, if
the network contains many cycles of varying length up to a maximum
$d$, the Fibonacci sequence generalizes to: $a_i = a_{i-1} + a_{i-2} +
\dots + a_{i-1-d} + a_{i-d}$, and the branching ratio satisfies: $d =
- \frac{\log (2-\varphi_d)}{\log \varphi_d}$ \cite{gardner}.

\subsection {Multi-layer composite fibers}
\label{multilayer}

Building blocks can also be combined to make composite fibers, like
prime numbers or quarks can be combined to form natural numbers or
composite particles like protons and neutrons, respectively.  The
ability to assemble fiber building blocks to make larger composites is
important in that it helps to understand systematically higher order
functions of biological systems composed of many genetic elements.  We
discover a particular type of composite made up of two elementary
building blocks, that we name multi-layer composite fiber. For
instance, the double layer {\it add-oxyS} fiber in the {\it crp-fis}
connected component (see Figs.  \ref{components}a and \ref{blocks}b
bottom, and ID\# 7 in SI Table \ref{all_building_blocks} and
Supplementary File 1) is a composite $|add-oxyS\rangle = |0,1\rangle
\oplus |1,1\rangle$ made of a series of genes composing a single fiber
of type $|0,1\rangle = | {\it add, dsbG, gor, grxA, hemH, oxyS, trxC}
\rangle$ that are regulated by two different transcription factors
       {\it rbsR} and {\it oxyR} that form another fiber of type
       $|1,1\rangle = | rbsR, oxyR \rangle $.  This composite is of
       importance since it allows for information to be shared between
       two genes, for instance {\it add} and {\it oxyS}, which are not
       directly connected (in this case, separated by a distant in the
       network of length two).

Composite fibers satisfy a simple engineering 'sum-rule': elementary
fibers are composed in series of fibers in a predefined order where
the first layer is represented by an entry fiber (carrying
transcription factors), and the last layer is formed by a terminator
fiber of output genes (encoding enzymes), as shown in
Fig. \ref{blocks}b, bottom.  This multi-layer composite fiber is
biologically significant because genes in the output layer synchronize
a genetic module that implement the same function even though the
genes in the module are not directly connected, and, indeed, can be at
far distances in the network. Such functionally related modules could
not be identified by modularity algorithms \cite{newman} which cluster
nodes in modules of highly connected nodes.

We find that composite fibers are dominant in eukaryotes (yeast,
mouse, human, see Section \ref{bio}). They resemble the building
blocks of multilayered deep neural networks where each subsequent gene
in the layer synchronizes despite the fact that nodes can be distant
in the network.  More generally, composite fibers with multiple layers
streamline the construction of larger aggregates of fibration building
blocks performing more complex function in a coordinated fashion.
These composite topologies complete the classification of input trees.

  \subsection {Fibration landscape across
    biological networks, species and system domains}
\label{bio}

To study the applicability of fibration symmetries across domains of
complex networks we have analyzed 373 publically available datasets
(SI Section \ref{datasets}). Full details of each network and results
can be accessed at
\url{https://docs.google.com/spreadsheets/d/1-RG5vR_EGNPqQcnJU8q3ky1OpWi3OjTh5Uo-Xa0PjOc}. The
codes to reproduce this analysis are at \url{github.com/makselab} (SI
Section \ref{algorithm1}) and the full datasets at
\url{kcorelab.org}. We analyze biological networks spanning from
transcriptional regulatory networks, metabolic networks, cellular
processes networks and signaling pathways, disease networks, and
neural networks. We span different species ranging from A. thaliana,
E. coli, B. subtilis, S. enterica (salmonella), M. tuberculosis,
D. melanogaster, S. cerevisiae (yeast), M. musculus (mouse) to
H. sapiens (human).  The topological fiber numbers $|n,\ell \rangle$
allow us to systematically classify fibers across the different
domains in a unifying way.  We find that fibration symmetries are
found across all biological processes and domains. The fiber
distributions for each type of biological network calculated by
summing over the studied species are displayed in
Fig. \ref{reduction}a and the fiber distributions for each species
calculated over the type of biological networks are shown in
Fig. \ref{reduction}b. Our analysis allows to investigate the specific
attributes and commonalities of the fiber building blocks inside and
across biological domains.  We find a varied set of fibers that
characterize the biological landscape.
Certain features of the fiber number distribution are visible in the
transcriptional networks in Fig. \ref{reduction}a. For instance, a
tail with $\ell$ is seen in the $n=0$ class as well as in the $n=1$
class.
Across species (Fig. \ref{reduction}b), bacteria like {\it E. coli} or
{\it B. subtilus} display a majority of $n=0$ building blocks, while
higher level organisms like yeast, mouse and human display a majority
of more complex building blocks like multi-layers and Fibonaccis.

To test the existence of symmetry fibrations across other domains we
extend our studies to complex networks beyond biology ranging from
social, infrastructure, internet, software, economic networks and
ecosystems (details of datasets in SI Section \ref{datasets}). Figure
\ref{reduction}c shows the obtained fiber distributions for each
domain.  A normalized comparison across domains is visualized in
Fig. \ref{reduction}d showing the cumulative number of fibers over all
domains and species per network size of 10$^4$ nodes. The results
support the applicability of the concept of symmetry fibration beyond
biology to describe the building blocks of networks across all
domains.

\subsection {Gene co-expression and synchronization via symmetry fibration}
 \label{coexpression}

We have shown in Section \ref{sync} that fibers in networks determine
cluster synchronization in the dynamical system. In a gene regulatory
network, symmetric genes in a fiber
synchronize their activity to produce gene co-expression levels that
sustain cellular functions.  We corroborate this result numerically in
Fig.~\ref{cpxr}g in the particular example of the {\it baeR-spy} FFF
in {\it E. coli}, and this result applies to all fibers, irrespective
of the dynamical system law.

To exemplify the synchronization in fibers, we consider the dynamics
in the composite fiber $|add-oxyS\rangle = |0,1\rangle \oplus
|1,1\rangle$ depicted in Fig. \ref{components}a and Fig. \ref{blocks}b
bottom, which is composed of autoregulator $1= crp$, and two layers of
fibers: $2= rbsR$, $3 = oxyR$, and $4 = add$, $5 = oxyS$ (we consider
here a reduced fiber for simplicity, and we add the autoregulator to
{\it crp} to the building block for completeness).
Graph $G = \{N_G, E_G\}$ consists of $N_G = \{1, 2, 3, 4, 5\}$, $E_G =
\{1 \rightarrow 1, 1 \rightarrow 2, 1 \rightarrow 3, 2 \dashv 2, 3
\dashv 3, 2 \rightarrow 4, 3 \rightarrow 5\}$ ($\dashv$ refers to
repressor and $\rightarrow$ to activation) and a 5-dimensional total
phase space $PG = \mathbb{R}^5$ with state vector $X(t) = \{x_1(t),
x_2(t), x_3(t), x_4(t), x_5(t)\}$ describing the expression levels of
each gene's product (e.g., mRNA concentration).

The symmetry fibration $\psi: G \to B$ collapses the graph $G$ into
the base $B = \{N_B, G_B\}$, where $N_B = \{a, b, c\}$ and $E_B = \{a
\rightarrow a, a \rightarrow b, b \dashv b, b \rightarrow c\}$. The
symmetry fibration acts on the nodes: $ \psi(1) = a$, $ \psi(2) =
\psi(3) = b$, $ \psi(4) = \psi(5) = c$, and on the edges: $\psi(1
\rightarrow 1) = a \rightarrow a$, $\psi(1 \rightarrow 2) = \psi(1
\rightarrow 3) = a \rightarrow b$, $\psi(2 \dashv 2) = \psi(3 \dashv
3) = b \dashv b$, and $\psi(2 \rightarrow 4) = \psi(3 \rightarrow 5) =
b \rightarrow c$.  Thus, the fibers partition the graph $G$ as $\Pi =
\{\Pi_a, \Pi_b, \Pi_c\}$, where $\Pi_a = \{1\}$, $\Pi_b = \{2, 3\}$
and $\Pi_c = \{4, 5\}$.

We represent the dynamics by two functions $k(x)$ and $g(x)$ modeling
degradation and synthesis of gene product, respectively
\cite{karlebach,klipp}. For example, $k(x)$ can be modeled as a linear
degradation term and $g(x)$ as a Hill function \cite{karlebach}.  We
consider that multiple inputs are combined by multiplying functions
$g(x)$, but any other way of combining inputs can be used.
Then, the dynamics of the expression levels of the genes in the
circuit are described by:

\begin{equation}
  \begin{cases}
    \frac{dx_1}{dt} = - k(x_1) + g(x_1) \\
    \frac{dx_2}{dt} = - k(x_2) + g(x_1) * g(x_2) \\
    \frac{dx_3}{dt} = - k(x_3) + g(x_1) * g(x_3) \\
    \frac{dx_4}{dt} = - k(x_4) + g(x_2) \\
    \frac{dx_5}{dt} = - k(x_5) + g(x_3) \, .\\
  \end{cases}
  \label{composite}
\end{equation}

The dynamics of the base are described by the state vector of the
base: $(y_a(t), y_b(t), y_c(t))$ with dynamical equations
\cite{sanders}:

\begin{equation}
  \begin{cases}
    \frac{dy_a}{dt} = - k(y_a) + g(y_a) \\
    \frac{dy_b}{dt} = - k(y_b) + g(y_a) * g(y_b) \\
    \frac{dy_c}{dt} = - k(y_c) + g(y_b) \, .\\
  \end{cases}
  \label{base}
\end{equation}

If $(y_a(t), y_b(t), y_c(t))$ is a solution for  the
base Eqs. (\ref{base}), then the map $\mathbb{P}_\psi$ sends the phase
space of this base to the phase space of the solutions in the graph
$G$ \cite{sanders}:

\begin{equation}
  \Big(x_1(t),x_2(t), x_3(t), x_4(t), x_5(t)\Big) = \mathbb{P}_\psi \Big[ y_a(t), y_b(t), y_c(t) \Big] = \Big(y_a(t), y_b(t), y_b(t), y_c(t), y_c(t) \Big) \, .
\end{equation}

Therefore, the graph $G$ sustains  a polysynchronous subspace (see for
instance Motivating example 1.4 in
\cite{deville}):

\begin{equation}
  \Delta_\Pi=\{(x_1, x_2, x_3, x_4, x_5) \in \mathbb{R}^5 \, \mid \, x_1(t), x_2(t) = x_3(t), x_4(t) = x_5(t) \} \, .
\end{equation}

This result can be corroborated by simply plugging $\Big(x_1(t),
x_2(t), x_3(t)=x_2(t), x_4(t), x_5(t)=x_4(t)\Big)$ into
Eqs. (\ref{composite}) to obtain a solution of the dynamics, implying
the synchrony $x_2(t)=x_3(t)$ in fiber $\Pi_b$ and $x_4(t)=x_5(t)$ in
fiber $\Pi_c$. We note that the concept of sheaves and stacks might be
useful to generalize the symmetry fibration framework to multiplex
networks.


We test this gene synchronization with publically available
transcription profile experiments available from the literature. We
use gene expression data profiles in {\it E. coli} compiled at Ecomics
\url{http://prokaryomics.com} \cite{ecomics}.  This portal collects
microarray and RNA-seq experiments from different sources such as the
NCBI Gene Expression Omnibus (GEO) public database~\cite{geo} and
ArrayExpress~\cite{array} under different experimental growth
conditions. The data is also compiled at the Colombos web
portal~\cite{colombos}.  The database contains transcriptome
experiments measuring the expression level of 4,096 genes in {\it
  E. coli} strains over 3,579 experimental conditions which are
described as: strain, medium, stress, and perturbation.  Raw data is
pre-processed to obtain expression levels by using noise reduction and
bias correction to normalize data across different platforms
\cite{ecomics}.

{\it E. coli} can adapt its growth to the different conditions that
finds in the medium. This adaptation is made by sensing extra and
intracellular molecules and using them as effectors to activate or
repress transcription factors. This implies that the different fibers
are activated by specific experimental conditions.  The Ecomics portal
allows to obtain those experimental conditions where a set of genes
has been significantly expressed under a particular set of
conditions. We perform standard gene expression analysis (see
\url{colombos.net} and Ref. \cite{colombos}) of the expression levels
in {\it E. coli} obtained under these conditions.

For a given set of genes in a fiber, we find the experimental
conditions for which the genes have been significantly expressed by
comparing the expression samples over the 4,096 different growth
conditions. Following \cite{colombos}, the experimental conditions are
ranked with the inverse coefficient of variation (ICV) defined as
$\mbox{ICV}_k=|\mu_k|/\sigma_k$, where $\mu_k$ is the average
expression level of the genes in the condition $k$ and $\sigma_k$ is
the standard deviation.  Following \cite{colombos}, we select those
conditions with ICV$_k>1$, i.e., where the average expression levels
in the particular condition $k$ are higher than the standard
deviation.  This score reflects the fact that, in a relevant
condition, the genes show an increment of their expression above the
individual variations caused by random noise.  Details on the
expression analysis can be found at Ref. \cite{colombos} and
\url{https://doi.org/10.1371/journal.pone.0020938.s001}.  Thus, we
obtain expression levels organized by the relevant experimental
conditions which are labeled according to the GEO database \cite{geo}.
From these data, we calculate the co-expression matrix using the
Pearson correlation coefficient between the expression levels of two
genes $i$ and $j$ in the relevant conditions for genes in a fiber. For
off-diagonal correlations between genes in different fibers, we use
the combined sets of conditions of both genes.


Results for the correlation matrix are shown in Fig. \ref{components}a
(bottom) for fibers regulated by the {\it crp-fis} strongly connected
component. Gene expression is obtained for every gene, so we plot the
correlation matrix calculated over each pair of genes. Genes that
belong to the same operon are transcribed as a single unit by the same
mRNA molecule, so these genes are expected to trivially synchronize
(variations exist due to attenuators inside the operon).  Thus, we
group together these genes as operons in the figure to indicate this
trivial synchronization.
To test the existence of fiber synchronization we compare gene
co-expression belonging to different operons. Figure \ref{components}a
(bottom) shows that expression levels of the genes that belong to a
fiber are highly correlated as predicted by the symmetry
fibration. Genes that belong to different fibers show no significant
correlations among them. In particular, there is no significant
correlation between the expression of genes in a given fiber and the
two master regulators {\it crp} and {\it fis}. This result is
consistent with the fibration symmetry and occurs despite the fact
that both, {\it crp} and {\it fis}, directly regulates all genes in
the studied fibers. We find some off-diagonal weak correlations
between fibers (e.g., {\it malI}), probably indicating missing links
or missing regulatory processes that produce extra
synchronizations. Some genes present weak correlations inside fibers
(e.g., {\it cirA}), indicating weak symmetry breaking probably from
asymmetries in the strength of binding rate of transcription factors
or input functions; effects that are not considered in the topological
view of the input trees, and can lead to desynchronization inside the
fiber.

\section{Discussion}

Fibration symmetries make sure that genes are turned on and off at the
right amount to assure the synchronization of expression levels in the
fiber needed to execute cellular functions.  In the fibration
framework, network function can be pictured as an orchestra in which
each instrument is a gene in the network. When the instruments play
coherently, with structured temporal patterns, the network is
functional. Here we have concentrated on the simplest temporal
organization, one in which some units (instruments) act synchronously
in time, a ubiquitous pattern observed in all biological networks. Our
findings identify the symmetries that predict this synchronization and
give rise to functionally related genes from the fibrations of the
genetic network.

Unlike network motifs which are identified by statistical
overrepresentation \cite{alon}, fibers in biology arise from
principles of symmetries following the tradition of how the building
blocks of elementary particles have being discovered in physics and
geometry \cite{weinberg}.  Our first principle approach to identify
building blocks is based on the circuit's theoretical and practical
(rather than statistical) significance to serve minimal forms of
coherent function and logic computation.

Further results shown in ~\cite{morone} indicate that
symmetries also describe the structure of neural connectomes and these
symmetries factorize according to function. Thus, symmetries can be
used to systematically organize biological diversity into building
blocks using invariances in the information flow encoded in the
topologies of the input trees. Genes related by symmetries are
co-expressed, thus providing a functional rationale for the biological
existence of these symmetries.

{\bf Acknowledgments} 

Research was sponsored by NIH-NIGMS R01EB022720, NIH-NCI
U54CA137788/U54CA132378, NSF-IIS 1515022 and NSF-DMR 1308235. We thank
L. Parra, W. Liebermeister, C. Ishida, M. S\'anchez and J. D. Farmer
for discussions. FM, IL, and HAM designed research, performed research, analyzed data, and wrote the paper.

\clearpage

FIG.~\ref{cpxr}. {\bf Definition of input tree, symmetry fibration,
  fiber and base}.  {\bf a,} The circuit controlled by the {\it cpxR}
gene regulates a series of fibers as shown by the different colored
genes.  The circuit regulates more genes represented by the dotted
lines which are not displayed for simplicity. The full lists of genes
and operons in this circuit are in SI Table \ref{all_building_blocks},
ID=27, 28 and 54.
{\bf b,} The input tree of representative genes involved in the {\it
  cpxR} circuit showing the isomorphisms that define the fibers.  For
each fiber, we show the number of paths of length $i-1$ at every layer
of the input tree, $a_i$, and its branching ratio $n$.
{\bf c,} Isomorphism between the input trees of {\it baeR} and {\it
  spy}.  The input trees are composed of an infinite number of layers
due to the autoregulation loop at {\it baeR} and {\it cpxR}. How to
prove the equivalence of two input trees when they have an infinite
number of levels?  A theorem proven by Norris ~\cite{norris}
demonstrates that it suffices to find an isomorphism up to $N-1$
levels, where $N$ is the number of nodes in the circuit. Thus, in this
case, 2 levels are sufficient to prove the isomorphism.
{\bf d, } Symmetry fibration $\psi$ transforms the {\it cpxR} circuit
$G$ into its base $B$ by collapsing the genes in the fibers into one.
{\bf e,} Symmetry fibration of the {\it fadR} circuit and {\bf f,} its
isomorphic input trees. Full list of genes in this circuit appears in
SI Table \ref{all_building_blocks}, ID=3, 4, and 58.  {\bf g,}
Symmetric genes in the fiber synchronize their activity to produce
same activity levels.  We use the mathematical model of gene
regulatory kinetics from Ref. \cite{alon-ecoli} (sigmoidal
interactions lead to qualitatively similar results) to show the
synchronization inside the fiber {\it baeR-spy} when the fiber is
activated by its regulator {\it cpxR}. Notice that {\it cpxR} does not
synchronize with the fiber.


FIG. \ref{components}. {\bf Strongly connected components of the
  genetic network and synchronization of gene co-expression in the
  fibers in {\it E. coli}}. {\bf a, Top,} Two-gene connected component
of {\it crp-fis}. This component controls a rich set of fibers as
shown. We also show the symmetry fibration collapsing the graph to the
base. We highlight the fiber {\it uxuR-lgoR} which sends information
to its regulator {\it exuR} and forms a 2-Fibonacci fiber $\rvert
\varphi_2 = 1.6180.., \ell = 2 \rangle$, as well as the double-layer
composite $|add-oxyS\rangle = |0,1\rangle \oplus |1,1\rangle$.  {\bf
  a, Bottom.}  Co-expression correlation matrix calculated from the
Pearson coefficient between the expression levels of each pair of
genes in Fig. \ref{components}a. Synchronization of the genes in the
respective fibers is corroborated as the block structure of the
matrix.
{\bf b,} The core of the {\it E. coli} network is the strongly
connected component formed by genes involved in the pH system as
shown.  This component supports two Fibonacci fibers: 3-FF and 4-FF
and fibers as shown.  Hollow colored circles indicate genes that are
in fibers and also belong to the pH component.


FIG.~\ref{blocks}. {\bf Classification of building blocks in {\it
    E. coli}}.  {\bf a, Basic fiber building blocks}. These building
blocks are characterized by a fiber that does not send back
information to its regulator. They are characterized by two integer
fiber numbers: $|n, \ell\rangle$. We show selected examples of
circuits and input trees and bases. The full list of fibers appears in
SI Table \ref{all_building_blocks} and Supplementary File 1.  The
statistical count of every class is in SI Table~\ref{statistics}.  The
last example shows a generic building block for a general n-ary tree
$|n, \ell\rangle$ with $\ell$ regulators.
 {\bf b, Complex Fibonacci and multilayer building blocks}.  These
 building blocks are more complex and characterized by an
 autoregulated fiber that sends back information to its
 regulator. This creates a fractal input tree that encodes a Fibonacci
 sequence with golden branching ratio in the number of paths $a_i$
 versus path length, $i-1$. When the information is sent to the
 connected component that includes the regulator, then a cycle of
 length $d$ is formed and the topology is a generalized Fibonacci
 block with golden ratio $\varphi_d$ as indicated. We find three such
 building blocks: 2-FF, 3-FF and 4-FF.  Last panel shows a multilayer
 composite fiber with a feed-forward structure.

 FIG. \ref{reduction}.  {\bf Fibration landscape across domains and
   species}. {\bf a, Fibration landscape for biological networks}. Total
 number of fiber building blocks across 5 types of biological networks
 analyzed in the present work.  The count includes the total number of
 fibers in the networks of each biological type considering all
 species analyzed for each type (see SI Table \ref{networks}).  {\bf
   b, Fibration landscape across species}. Count of fibers across each
 analyzed species. Each panel shows the count over the different type
 of biological networks ({\it E. coli} contains only the
 transcriptional network, see SI Table \ref{networks}).  {\bf c, Fibration
   landscape across domains}. Count of fibers across the major domains
 studied. The biological domain panel is calculated over all networks
 and species in {\bf a} and {\bf b}.  {\bf d, Global fibration
   landscape}. Cumulative count of fibers in all domains in {\bf c}.
 The cumulative count represents the total number of fibers per
 network of $10^4$ nodes.  Specifically, the quantity is calculated as
 the total number of fibers divided by the total number of nodes in
 all networks per domain multiplied by $10^4$.

\clearpage

\begin{figure*}[h!]
    \includegraphics[width=.95\textwidth]{Fig1.pdf}
\caption {}
\label{cpxr}
\end{figure*}

\clearpage

\begin{figure*}
  \includegraphics[width=.95\textwidth]{Fig2a.pdf}
  \centering
  \caption{{\bf a}}
\label{components}
\end{figure*}
\clearpage

\addtocounter{figure}{-1}
\begin{figure*}
    \includegraphics[width=.8\textwidth]{Fig2b.pdf}
    \caption{{\bf b}}
\end{figure*}


\clearpage

\begin{figure*}[h]
\includegraphics[width=.72\textwidth]{Fig3a.pdf} 
\centering
\caption {{\bf a}}
\label{blocks}
\end{figure*}

\clearpage
\addtocounter{figure}{-1}
\begin{figure*}[h]
\includegraphics[width=.7\textwidth]{Fig3b.pdf} 
\centering
\caption {{\bf b}}
\end{figure*}

\clearpage

\begin{figure}[h]
\includegraphics[width=.95\textwidth]{Fig4ab.pdf}
\caption{{\bf a, b}}
\label{reduction} 
\end{figure}

\clearpage
\addtocounter{figure}{-1}
\begin{figure}[h]
\includegraphics[width=\textwidth]{Fig4cd.pdf}
\caption{{\bf c, d}}
\end{figure}


\clearpage

\centerline{ \bf Supplementary Information}

\centerline{Fibration symmetries uncover the building blocks of biological
  networks}

\medskip

\centerline{ \bf Flaviano Morone, Ian Leifer, Hern\'an A. Makse}

\tableofcontents

\clearpage

\section{Transcriptional regulatory network of {\it E. coli}}
\label{transcriptional}

To define the transcriptional regulatory network (TRN) we use the
transcription factor-gene target bi-partite network of {\it
  Escherichia coli} K-12 obtained from the RegulonDB data source
(\url{http://regulondb.ccg.unam.mx}). RegulonDB manually curates all
transcriptional regulations from literature searches
~\cite{regulon}. We download all transcriptional regulatory
interactions catalogued in RegulonDB version 9.0 from
\url{http://regulondb.ccg.unam.mx/menu/download/datasets/files/network_tf_gene.txt},
last accessed September 15, 2018.

The database downloaded from RegulonDB is composed of a bipartite
transcription factor - gene target network. In this bi-partite
dataset, a directed link between a source transcription factor (TF)
and a target gene means that the TF binds to the DNA sequence at the
binding site of the target gene to regulate its rate of transcription.
In {\it E. coli}, each gene expresses a single TF (this is not the
case in eukaryotic genes that contains introns and splicing of
protein-coding RNA can produce many proteins from a single
gene). Therefore, a gene-gene regulatory network can be constructed
from the bipartite transcription factor-gene target network by
associating each TF to the gene that expresses the TF. Then, a
directed link in the TRN from gene $i$ $\to$ gene $j$ implies that
gene $i$ encodes for a TF that controls the rate of transcription of
gene $j$.  Thus, a directed link encodes the combined processes of
transcription, translation and TF binding to a target gene.  We denote
genes in bacteria in italics, e.g., {\it gadX} and its protein as
GadX.  Thus, we say that gene $i$ sends a genetic 'message' to gene
$j$ and the 'messenger' is the TF. The history of all messages passing
in the network defines the information flow in the network.  A TF can
either be an activator, repressor or can have a dual function. For the
purpose of calculating isomorphisms between input trees, the dual
interactions are treated as distinct interactions.  Thus, these three
interactions are treated as three different types.

For the purpose of building the TRN it is important to distinguish the
gene's products between genes encoding for TFs and the rest of the
genes encoding for the rest of the proteins (enzymes, kinases,
transport proteins, etc). A TF is a regulatory protein that regulates
a gene by binding, and therefore will always have an out-going link in
the network. There are other regulatory proteins (like kinases,
histones, coactivators, etc) that regulate gene expression but they do
not have a DNA-binding domain and they regulate gene expression
without binding. In our TRN, genes that encode for a protein that is
not a TF do not have out-going links in the network. They only have
in-going links and therefore are dangling ends in the network. In {\it
  E. coli} most of these proteins are enzymes that catalyze
biochemical reactions in the metabolic network. Other proteins are
involved in transport and signaling processes (kinase) in the cell.

TF are also activated by effector molecules (metabolites) that bind
non-covalently to an allosteric site of the TF to alter the
conformation of the TF to activate it or deactivated by controlling
the binding/unbinding of the TF to DNA. Effectors can also produce
covalent activation of the TF like for instance during phosphorylation
mediated by kinases in the two component TFs.

We treat these effector activities as external parameters, determined
by the growth conditions in the surrounding system (the cell in its
changing environment) or by the metabolic network, which is considered
external to the TRN. These external perturbations are considered as
the external growth conditions when we analyze the co-expression
profiles in Section \ref{coexpression}.  In the present study, the
metabolic network is considered external to the TRN, so we do not
consider feedback loops from the TRN to the metabolic network and back
to the TRN mediated by effector metabolites. This extended network is
treated in a follow up.

In {\it E. coli}, genes are also grouped by operons. An operon is a
set of contiguous genes that are transcribed as a single unit from the
same mRNA molecule and the same promoter site upstream of all genes
and a terminator downstream \cite{regulon}. An operon can contain
genes encoding for TF or non-TF proteins, and more than two TFs can be
part of the operon. Since the operons are transcribed by the same RNA
molecule, then we group these genes into a single node in the
network. This is certainly the case when the operon has a single
promoter transcribing the full operon.  However, there is some
ambiguity in the construction of the network using the definition of
operon in RegulonDB when there are promoters in the middle of the
operon and these promoters transcribe more than one TF in the operon,
forming different transcription units.  For instance, the operon in
the gad system, {\it gadAXW} which is important in the pH strongly
connected component in Fig. \ref{components}b. This operon expressed
two TFs, GadX and GadW, and one enzyme GadA. Here, each gene has its
own promoter and terminator and thus are different nodes in the
network. Moreover, each TF is regulated by different TFs as well as
each TF regulates different genes.  As seen in Fig. \ref{components}b,
for instance, GadX binds to {\it hns} but not GadW. Also, GadW is
regulated by {\it ydeO} but {\it ydeO} does not regulate {\it
  gadX}. Thus, putting together these two genes in the same operon
{\it gadAXW} would miss all these links. Thus, when two TF with
different promoters are part of the operon, we consider the TF as
different genes. On the other hand, the non-TF genes in operons are
always put together with other genes in the operon.  For instance, the
{\it gadAXW} operon from RegulonDB is considered as two nodes: {\it
  gadW} and {\it gadAX}. To simplify notation, when there is an operon
that contains one TF and several non-TF proteins, then for simplicity,
we call this operon by the name of the TF. For instance, {\it gadAX}
is simply called {\it gadX} or the operon {\it rbsDACBKR} is called
{\it rbsR} and therefore the TF {\it rsbR} represents the entire
operon {\it rbsDACBKR}. Finally, when all the genes in the operon are
non-TF, then we call the operon with all the genes names, as for
instance, {\it lsrACDBFG-tam}.

In the RegulonDB database there are a total of 4690 genes.  Out of
these genes, Regulon DB provides a bipartite network consisting of
1843 genes with interactions from or to other genes, the remaining
genes are not considered in the analysis. There are 192 genes that
encode for TFs.  We cluster the genes into 313 operons as explained
above.  Full names of operons and genes appear in SI Table
\ref{all_building_blocks}.  After grouping the genes into operons, the
network is reduced to 879 nodes. There are 1835 directed edges with an
average in-degree (or out-degree) of 2.1.  In this network we find 91
different fibers that encompass 416 different nodes. We find that 28
nodes are involved in 7 strongly connected components of size larger
than one node, and the rest are single node connected components.

\begin{figure*}[h!]
\includegraphics[width=.95\textwidth]{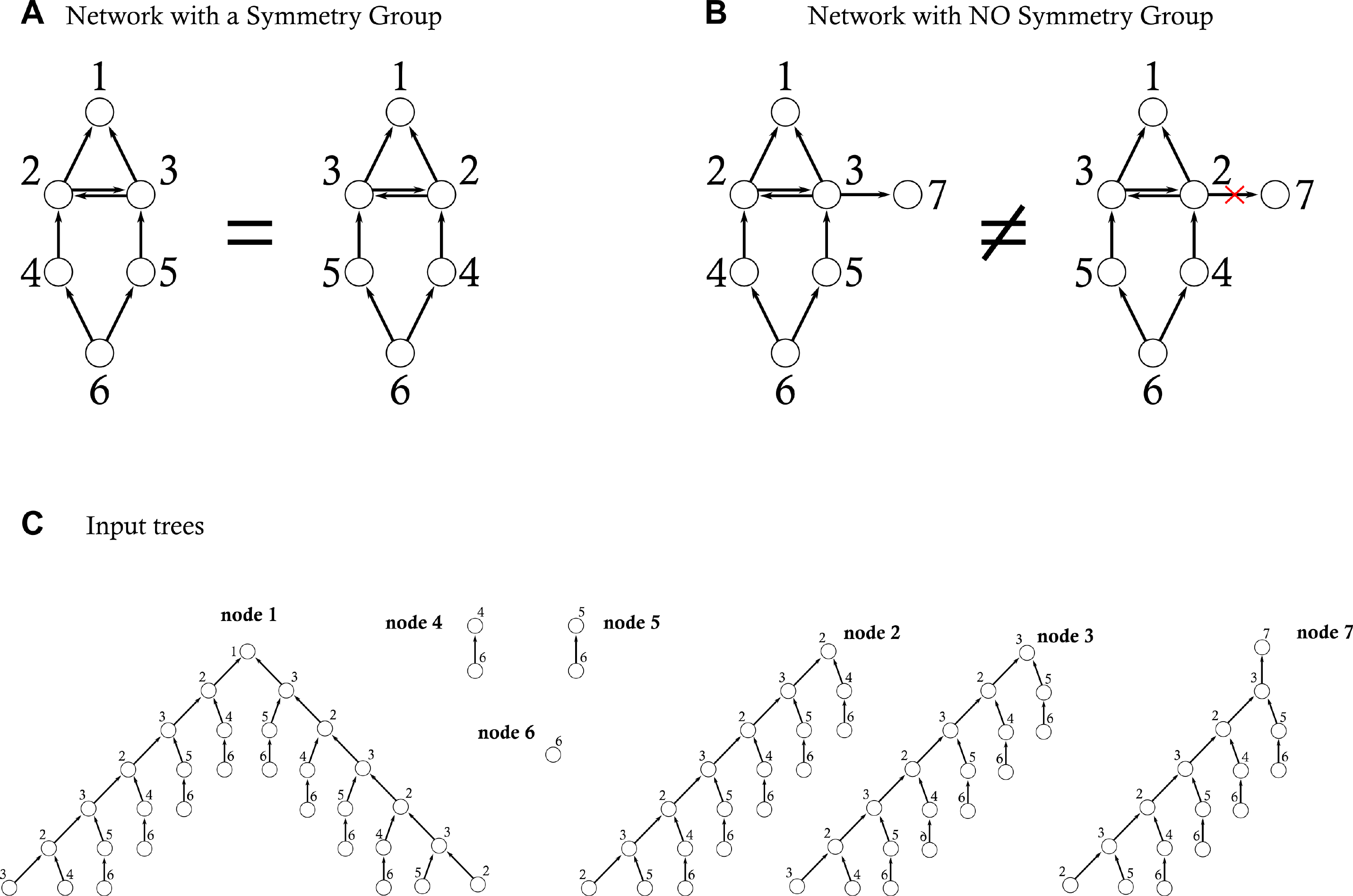} \centering
\caption {{\bf Group symmetries and fibrations with their input tree}.
  {\bf a, } Example of a network with a symmetry group.  The
  automorphism shown maps the network into another network leaving
  invariant the connectivity of every nodes in the network
  \cite{dixon,golubitsky,other0,pecora1}. {\bf b,} A network without
  automorphisms but with a fibration.  The addition of a single
  out-link from 3 $\to 7$ breaks the whole group symmetry.  However,
  since fibrations are defined according only to the input tree, then
  the network still have a symmetry, a fibration arising from the fact
  that the input trees of nodes 2 and 3 are isomorphic, as well as
  between the input trees of nodes 4 and 5 as shown in {\bf (c).}
  There are no more isomorphisms as shown by the rest of the
  input trees. Therefore, nodes 2 and 3 form a fiber.
  Nodes 4 and 5 also form another fiber,
  yet independently of the other fiber. The fibration is a morphism
  that maps the network into a base which is formed by collapsing the
  isomorphic nodes into one, i.e., collapsing node 2 and 3 together,
  and node 4 and 5 together. The resulting base is also called a
  quotient graph.}
\label{example} 
\end{figure*}


\section{Symmetry fibrations}
\label{sec:fibration}





Below we provide formal definitions of the main concepts using in the
paper: (a) input trees and isomorphisms, (b) from fibrations $\to$
surjective minimal graph fibrations called here symmetry fibrations,
(c) fibers and minimal bases, and (d) minimal balance coloring
algorithm.  We start with a review of the literature (not exhaustive).

The literature on fibrations and groupoids crosses the fields of
mathematics, computer science and dynamical systems theory. The notion
of fibration was first introduced by Grothendieck as fibrations
between categories in algebraic geometry ~\cite{grothendieck}.  The
original paper of Grothendieck has been published as a part of the
S\'eminaire N. Bourbaki in 1958 and can be found at
~\url{http://www.numdam.org/article/SB_1958-1960__5__299_0.pdf}. A
mathematical account of Grothendieck fibrations in the context of
category theory appears in
\url{https://ncatlab.org/nlab/show/Grothendieck+fibration}. For a
review of the history of fibrations from Grothendieck to modern
studies, see the blog of Vigna at
~\url{http://vigna.di.unimi.it/fibrations/}. The formulation of
Grothendieck is highly abstract and differs from our present work
which refers to the notion of surjective minimal graph fibration which
is a fibration between graphs.  The work of Boldi \& Vigna
\cite{boldi} and DeVille \& Lerman ~\cite{deville} on graph fibrations
are the closest to our formulation, see
~\url{http://vigna.di.unimi.it/ftp/papers/FibrationsOfGraphs.pdf}.
Graph fibrations have been applied in computer science to understand
PageRank~\cite{boldi-pagerank}, and the state of synchrony of
processors in computing distributed
systems~\cite{boldi-distributed,boldi-sync}, where fibrations are the
key concept in the computation of identical states in distributed
system.  The relation between surjective minimal graph fibrations and
synchronous subspaces is elaborated in DeVille \& Lerman
~\cite{deville} and Nijholt, Rink \& Sanders \cite{sanders}.  It
should be noted that all these works on fibrations pertain to a highly
abstract mathematical level which, in turn, provides the concept of
fibration with a quite broad applicability. For a more accessible
reading on fibrations within the particular context application to
biological networks, the reader is recommended to follow our paper and
supplementary sections.

In parallel, the work of Golubitsky and
Stewart~\cite{golubitsky,stewart} and others in dynamical systems
theory consider the equivalent formalism of symmetry groupoids,
equitable partition of balanced colored nodes and its relation with
synchronization \cite{arenas,kurths,strogatz}. A review of the
groupoid formalism and its application to synchronization in dynamical
systems appears in~\cite{golubitsky}.
DeVille and Lerman~\cite{deville} also discuss the relation between
graph fibrations and the groupoid formalism.

Synchronization arises also as a consequence of permutation
symmetries in the network, called automorphisms \cite{dixon}, which
form symmetry groups and are different from symmetry fibrations and
symmetry groupoids. There is a large literature in the dynamical
system community dealing with cluster synchronization from
automorphisms, since synchronization is an ubiquitous phenomenon
across all sciences \cite{arenas,kurths,strogatz}. Reviews can be
found in the work of Golubitsky and Stewart \cite{golubitsky,stewart}
to recent work in ~\cite{other0,pecora1,pecora2} and references
therein. Symmetry groups are the cornerstone of physical phenomena
appearing in all physical systems \cite{weinberg}.

Below, to elaborate on the definition of symmetry fibrations, we first compare
fibrations to automorphisms which form symmetry groups
\cite{dixon,golubitsky,other0,pecora1,pecora2}
using the example networks of Figs.~\ref{example}a and~\ref{example}b.
An automorphism is a transformation that preserves the full
connectivity of the network. That is, an automorphism preserves not
only the inputs but also the outputs of each node in the network, and
therefore, it presents more stringent conditions on the connectivity
than symmetry fibrations which preserve only the input trees.  For
example, the network of Fig.~\ref{example}a is invariant under the
automorphism defined by the permutation:
\begin{equation}
\sigma\ =\ \left(
\begin{matrix}
    1 & 2 & 3 & 4 & 5 & 6 \\
    \down & \down & \down & \down & \down & \down \\
    1 & 3 & 2 & 5 & 4 & 6 \\  
\end{matrix}
\right)\ ,
\label{eq:legal_perm}
\end{equation}
because the nodes are connected exactly to the same nodes before and
after the application of the permutation $\sigma$, which is a global
mirror symmetry.

Next, consider the slightly modified network depicted in
Fig.~\ref{example}b left, which differs from the network in
Fig.~\ref{example}a by one extra out-going link from node 3 to 7. In
this network, the permutation of nodes $2\leftrightarrow3$ and
$4\leftrightarrow5$, Eq. (\ref{eq:legal_perm}), is not an automorphism
anymore, because it does not preserve the in and out connectivities of
all nodes, e.g., node 3 is connected with 7 but loses this connection
after the permutation (Fig.~\ref{example}b right). It is interesting
to see how fragile group symmetries are: if we connect just one extra
node to the network as shown in Fig.~\ref{example}b, the symmetry
(i.e. the network automorphism group) is broken.  This occurs because
automorphisms
require very strict arrangements of nodes and links to preserve,
rigidly, the global structure of the network.  Fibration symmetries,
with their emphasis in the preservation of the input trees only, is
less restrictive. This might explain why fibration symmetries emerged
in living systems as opposed to the more restrictive automorphisms
which describe all aspects of matter, from elementary particles to
atoms, molecules and phases of matter.

This example raises the following question: are there extra symmetries
in the network shown in Fig.~\ref{example}b beyond 
its automorphisms? The answer to this question is, indeed, yes: there
are extra symmetries in the network of Fig.~\ref{example}b, the
fibration symmetries \cite{grothendieck,boldi}, which do not form a
group \cite{dixon} but groupoids \cite{golubitsky}. A groupoid is a
set of transformations satisfying the axioms of invertibility,
identity and associativity but not the composition law (closure)
\cite{golubitsky}, while in a group, transformations satisfy the four
axioms. For this reason, groupoids are fundamentally different
algebraic structures compared with traditional group symmetries.



\subsection{ Input tree}
\label{sec:input}

Roughly speaking, symmetry fibrations take into account only the input
trees of the nodes, but not the output-trees (this is not true though
when the input and output trees are connected). Thus, node 3 in
Fig. \ref{example}b is connected to node 7 via an out-going link, and
this link destroys the symmetry group, but node 3 is still symmetric
with 2 via a symmetry fibration,
since the input trees of nodes 2 and 3 are isomorphic, even though
node 3 is connected with 7. This is because the connection $3\to 7$ is
an out-going link of node 3 and, therefore, is not part of its input
tree.  Simply put, symmetry fibrations preserve input trees only,
while automorphisms preserve both input and output-trees, since they
preserve the full connectivity of the network, and thus, they
represent more stringent symmetries than fibrations. We formalize this
idea next after introducing some definitions.

\bigskip

The basic ingredient to define 
a new symmetry beyond automorphisms is the {\bf input tree}, which
contains the full information received by a given node through the
totality of all the possible paths ending in that node and starting
from every other node in the network.
Thus, for every node $i$ in the network $G$ there is a corresponding
input tree, called $T_i$, which is defined as a tree with a selected
node $r_i$, called the root, and such that every other node is a path
$\mathcal{P}_{j\to i}$ of $G$ starting from $j$ and ending in
$i$~\cite{sanders}.  A link from node $\mathcal{P}_{j\to i}$ to node
$\mathcal{P}_{k\to i}$ exists if $\mathcal{P}_{j\to i} = e_{j\to
  k}\mathcal{P}_{k\to i}=$, where $e_{j\to k}$ is an edge of $G$.

The concept of input tree has appeared in the literature as the
universal total space in traditional categorical or topological
terminology~\cite{grothendieck}, the universal total graph
from~\cite{boldi}, the view in the theory of distributed systems, or
the unfolding of a nondeterministic automaton in concurrency
theory~\cite{boldi}.

For example, let us construct the input tree $T_2$ of node 2 in the
network on the left of Fig.~\ref{example}b. The root is the node $r_2$
at the uppermost level of the tree. Every other node of the input tree
of node 2 is a path $\mathcal{P}_{j\to 2}$ ending in 2. There are two
paths of length 1: $\mathcal{P}^{(1)}_{3\to 2}$ and
$\mathcal{P}^{(1)}_{4\to 2}$; three paths of length 2:
$\mathcal{P}^{(2)}_{2\to 2}$, $\mathcal{P}^{(2)}_{5\to 2}$, and
$\mathcal{P}^{(2)}_{6\to 2}$; and so on.  Since
$\mathcal{P}^{(2)}_{2\to 2}=e_{2\to 3}\mathcal{P}^{(1)}_{3\to 2}$, we
put a link in the input tree from $\mathcal{P}^{(2)}_{2\to 2}$ to
$\mathcal{P}^{(1)}_{3\to 2}$ because $\mathcal{P}^{(2)}_{2\to
  2}=e_{2\to 3}\mathcal{P}^{(1)}_{3\to 2}$. We then add all other
links in the input tree using the same criterion.  The resulting
input tree $T_2$ is shown in Fig.~\ref{example}c, together with the
input trees of all other nodes in the network in Fig.~\ref{example}b.

To simplify, we label each node of $T_i$ using the starting point of
the corresponding path $\mathcal{P}_{j\to i}$. For example, in $T_2$
nodes $\mathcal{P}^{(1)}_{3\to 2}$ and $\mathcal{P}^{(1)}_{4\to 2}$
are labeled 3 and 4 respectively, and the length of the path is equal
to the depth of the node in the input tree.

Thus, in practice, we arrive at the following way to construct the
input tree: we start with the node at the root, lets say node 2. We
label every node $\mathcal{P}_{j\to 2}$ in the input tree by node $j$
where the path starts. The first layer of the input tree consists of
all the nodes that are at a distance one from the root. In this case,
nodes 3 and 4. Thus we add two links to 2 from 3 and 4 in the input tree.

The second layer of the input tree is obtained applying the same
procedure to each node in the first layer, 3 and 4. For instance, node
3 receives a link from 2 and 5. Therefore the second layer of the
input tree contains nodes 2 and 5 connected to node 3. We repeat the
procedure with the other node in layer 2: node 4. Node 4 receives a
link only from node 6, and node 6 from no one. So, we add a link from
6 to 4 and this path does not propagate further. The third layer of
the input tree is obtained iteratively applying the same procedure,
and so on.

We note that the input trees of nodes 1, 2, 3 and 7 are infinite since
the network contains a cycle (or loop) between nodes
$2\rightleftarrows 3$.  For instance, $T_1$ is infinite because there
are paths crossing the loop infinite times.  On the other hand, the
input trees of nodes 4, 5 and 6 are finite since they do not cross the
loop.

\subsection{Isomorphic input trees}
\label{sec:isomorphic}

The input tree $T_i$ 
at node $i$ can be interpreted as the collection of all possible
`histories' starting at some node and ending in node $i$. As shown in
Section \ref{sync}, if two input trees $T_i$ and $T_j$ are isomorphic,
then the corresponding nodes $i$ and $j$ in network $G$ have the same
dynamical state ~\cite{deville,sanders}.  This equivalence is
understood in terms of a local in-isomorphism
that maps nodes to nodes and links to links, so it formalizes the fact
that the dynamical interactions represented by a directed link from
gene to gene could be in principle different across genes, as long as
the links are the same (or similar, in case that the produced
synchronization is approximate) inside the fiber.

An isomorphism between $T_i$ and $T_j$ is defined as a bijective map
$\tau: T_i\to T_j$, which maps one-to-one the nodes and edges of $T_i$
to nodes and edges of $T_j$.

A minimal condition for the existence of an isomorphism between the
input trees is that the two input trees have the same number of nodes
(we could also add a condition of the same degree sequence).  Thus, it
is clear that there could be no isomorphism between the input trees of
nodes 2 and 4, since the former contains an infinite number of nodes
and the later just two. Thus, a minimal condition for an isomorphism
to exist is that it should be a mapping between two input trees with
the same number of nodes, since the mapping needs to be bijective,
i.e., with an inverse.  By inspection it is then clear that there is
an isomorphism between the input trees of nodes 4 and 5. This
isomorphism is the map $\tau_{4 \to 5}: T_4\to T_5$, and it is written
as a transformation following the notation:
\begin{equation}
\begin{aligned}
&\tau_{4\to 5} = \left(
\begin{matrix}
    4: & 6 \\
    \downarrow & \downarrow \\
    5: & 6 
\end{matrix}
\right)\ , 
\end{aligned}
\,\,\,\,\, \mbox{(isomorphism
    between input trees of nodes 4 and 5).}
\label{eq:tau_45}
\end{equation}
which maps the root of $T_4$ to the root of $T_5$ 
as $\tau_{4\to 5}(4) = 5$, and node $6\in T_4$ to node $6\in T_5$ 
as  $\tau_{4\to 5}(6) = 6$. 
The notation starts with the root of the tree and then we write nodes
in each level from top to bottom starting from left to right in each
level.  
In this particular example the links are of the same type, so there is
no need to specify the mapping between links in the isomorphism, but
in general the local equivalence require that nodes are map to nodes
and also links are mapped to the same type of link by the isomorphism.

The map in Eq.~\eqref{eq:tau_45} is one of the simplest isomorphism
since the input tree contains only one level.  In this particular
case, to see that nodes $T_4$ and $T_5$ are isomorphic, it is thus
enough to see that both nodes 4 and 5 connect to one and the same
node, which is node 6 in this case.  That is, both input trees of
nodes 4 and 5 are isomorphic because they are made up of just two
nodes and one edge, and this isomorphism implies that 4 and 5 receive
the same information. This is the simplest form of an isomorphism
between input trees. In this case, we say that node 4 and 5 have the
same {\it input-set}, which is an input tree of only one level, that
is the set of incoming links. The input-set is used in the groupoid
formalism in Ref.~\cite{golubitsky}.


Next, we consider the input trees of nodes 2 and 3.  By visual inspection,
both input trees have the same `shape'. 
However, these trees are infinite in the number of levels. How do we
decide if two input trees are isomorphic when they have an infinite
number of levels?  Remarkably, to determine if two input trees are
isomorphic, it suffices to check that they are isomorphic up to the
$N-1$ level, thanks to a theorem by Norris~\cite{norris}, where $N$ is
the total number of nodes in the network $G$.  This is an important
result that allows us to avoid to check an infinite number of
equivalences. Since $G$ has $|N_G|=7$, we use six levels in the input
trees to determine that there is an isomorphism between $T_2$ and
$T_3$ which corresponds to the following map:
\begin{equation}
\begin{aligned}
&\tau_{2\to 3} = \left(
\begin{matrix}
2: & 3 & 4 & 2 & 5 & 6 & 3 & 4 & 6 &\dots\\
\down & \down & \down & \down & \down & \down & \down & \down & \down & \down\\
3: & 2 & 5 & 3 & 4 & 6 & 2 & 5 & 6 &\dots
\end{matrix}
\right),
\end{aligned}
\, \mbox{(isomorphism
    between input trees of 2 and 3).}
\label{eq:tau_23}
\end{equation}
There are no other isomorphism between the other input trees.
Notice that $T_7$ is not isomorphic to $T_3$ and $T_2$ by just one
link to the root.




The existence of an isomorphism $\tau$ from the input tree of node $i$
to the input tree of node $j$ implies the synchronization of $x_i$ and
$x_j$ \cite{deville}. In the groupoid formalism of Golubitsky and
Stewart, it is said that two nodes are synchronized if their input-set
are synchronized, too \cite{golubitsky}.
Analogous work in dynamical systems shows that automorphisms in
networks
lead to synchronized nodes in orbits, see
~\cite{other0,pecora1,pecora2,stewart} and references therein.  The
orbit of a given node is obtained by applying all automorphisms of a
network to the node and the nodes in the orbit are synchronous.  The
synchronized orbits obtained from automorphisms are analogous to the
synchronized fibers obtained from symmetry fibrations. In general,
every orbit is also a fiber, but the opposite is not true, since a
fiber is not necessarily an orbit.

In our analysis of the {\it E. coli} network, we find some
automorphisms. Some of the star fibers with $n=0$ are also orbits of
the networks since they are invariant under permutation symmetries of
the symmetric group of order $n$, $S_n$.  But this is only when the
genes in the star have no out-going links.  As shown in the example of
Fig. \ref{example}, an out-going link in any of the star genes, will
destroy the automorphism, but not the fiber.  For this reason,
automorphisms are somehow more prevalent in undirected networks. For
instance, we have found that automorphisms describe the symmetries of
the gap junction connectome of {\it C. elegans}, which is composed all
of undirected links \cite{morone}. In the case of directed
 biological networks treated here, while automorphisms
could be of use to discover some synchronized nodes, the majority of
synchronization is due to symmetry fibrations, which are not described
by automorphisms.

\medskip

\subsection{ From fibrations to symmetry fibrations via isomorphic input trees and minimal bases}
\label{sec:base}

\medskip
A fibration is any morphism from a network $G=(N_G, E_G)$ to a base
$G=(N_G, E_G)$: $\psi: G \to B$ \cite{grothendieck}.  If a network
$G=(N_G, E_G)$ has at least one pair of isomorphic input trees, then
there exists a network $B=(N_B, E_B)$, called the {\bf base} of $G$,
such that $G$ can be `fibered' over $B$ by the graph fibration.  The
base $B$ is defined as follows:
\begin{itemize} 
\item a node $I\in N_B$ is a representative 
of the set of nodes $\{i\in N_G\}$ whose input trees are isomorphic;  
\item an edge $e_{I\to J}$ where $I, J \in E_B$ is defined as $e_{I\to
  J} = \sum_{i\in I}e_{i\to j}$, where $e_{i\to j}\in E_G$.
\end{itemize}
Having defined the base network $B$, we say that $G$ is fibered over
$B$ if there exists a surjective morphism $\psi: G\to B$, called
surjective graph fibration \cite{boldi}, that maps nodes
and edges of $G$ to nodes and edges of $B$ as: $\psi(i)=I$ for all
$i\in N_G$, and $\psi(e_{i\to j}) = e_{I\to J}$.  A surjective morphism
is a map between two sets (the domain and codomain) where each element
of the codomain (in this case $B$) is mapped to, at least, by one
element of the domain (in this case $G$).
%
%
%
The set of nodes $i\in N_G$ that are mapped to the same node $I\in N_B$,
and denoted by $\psi^{-1}(I)$, is called the fiber of $G$ over node
$I$.  We notice that all input trees of nodes which belong to the same
fiber are pairwise isomorphic.

In general a surjective graph fibration $\psi$ can map nodes with
isomorphic input trees to different bases, thus, the number of fibers
is not minimal.

A surjective graph fibration that maps all genes with isomorphic input
trees to a single common node in $B$ is called a surjective minimal
graph fibration in the sense of \cite{boldi}.  Such a minimal
fibration will generate then the minimal bases of the network and will
produce the largest collapse of nodes in fibers.  In this work we only
deal with surjective minimal graph fibrations and we call them
symmetry fibrations for short. 

In practice, a symmetry fibration maps $G$ to the minimal base $B$
(analogous to the quotient), that consists of the following steps:
{\it (i)} consider all the nodes in a fiber (which have isomorphic
input trees) and choose one as the representative $I$, {\it (ii)}
collapse the nodes in the fiber into one single node in $B$ and call
it by the name of the representative node $I$, {\it (iii)} for every
link of a node $j$ in $G$ directed to the node $I$ in $G$, add a link
in $B$ from $j$ to $I$. If the node $j$ belongs to the fiber, then the
corresponding link in $B$ is an autoregulation loop in $B$, {\it (iv)}
repeat for every fiber in $G$. When fibers belong to disjoint
components of the network, then they are considered as distinct
fibers.




\section{Algorithm to find fibers with minimal balance coloring}
\label{algorithm1}


The algorithm to partition the network into fibers is based on the
'minimal balanced coloring' algorithm developed by Cardon \&
Crochemore in Ref.~\cite{cardon}.  Here we follow a version developed
by Kamei \& Cock \cite{kamei} to construct a {\it minimal} balanced
coloring of a network, namely a coloring that employs the least
possible number of colors, which is associated with minimal graph
fibrations.  The algorithm's runtime scales as $O( |E_G|
\log_2|N_G|)$, which implies that it is essentially linear with the
network size, specially for sparse networks, and can be applied to
very large networks.

The theory of balance coloring is explained in Ref.~\cite{golubitsky}.
A balance coloring creates a partition of nodes of $G$ into disjoint
sets (corresponding to synchronous fibers) such that each node in one
set receives the same number of colors from nodes within other
sets~\cite{golubitsky,stewart}. A coloring of $G$ with this property
is the {\it balanced coloring} and represents an {\it equitable
  partition} of the network, see~\cite{golubitsky,stewart}.  The sets
identified by a {\it minimal balanced coloring} partitions the network
with minimal colors and corresponds to the fibers of $G$ identified by
minimal graph fibrations $\psi$~\cite{boldi,deville,golubitsky}.

Thus, we color nodes such that synchronous nodes in a fiber receive
the same colors from their synchronous nodes. As example, the genes
{\it baeR} and {\it spy} (Fig.~\ref{cpxr}a) have the same color and are
in the same fiber since they receive the same colors from their
neighbors: both {\it baeR} and {\it spy} receive one red color via the
activator link from one red node ({\it baeR} from itself and {\it spy}
from {\it baeR}) and one green activator link each from the green node
{\it cpxR}.

\begin{figure}[h!]
\includegraphics[width=\textwidth]{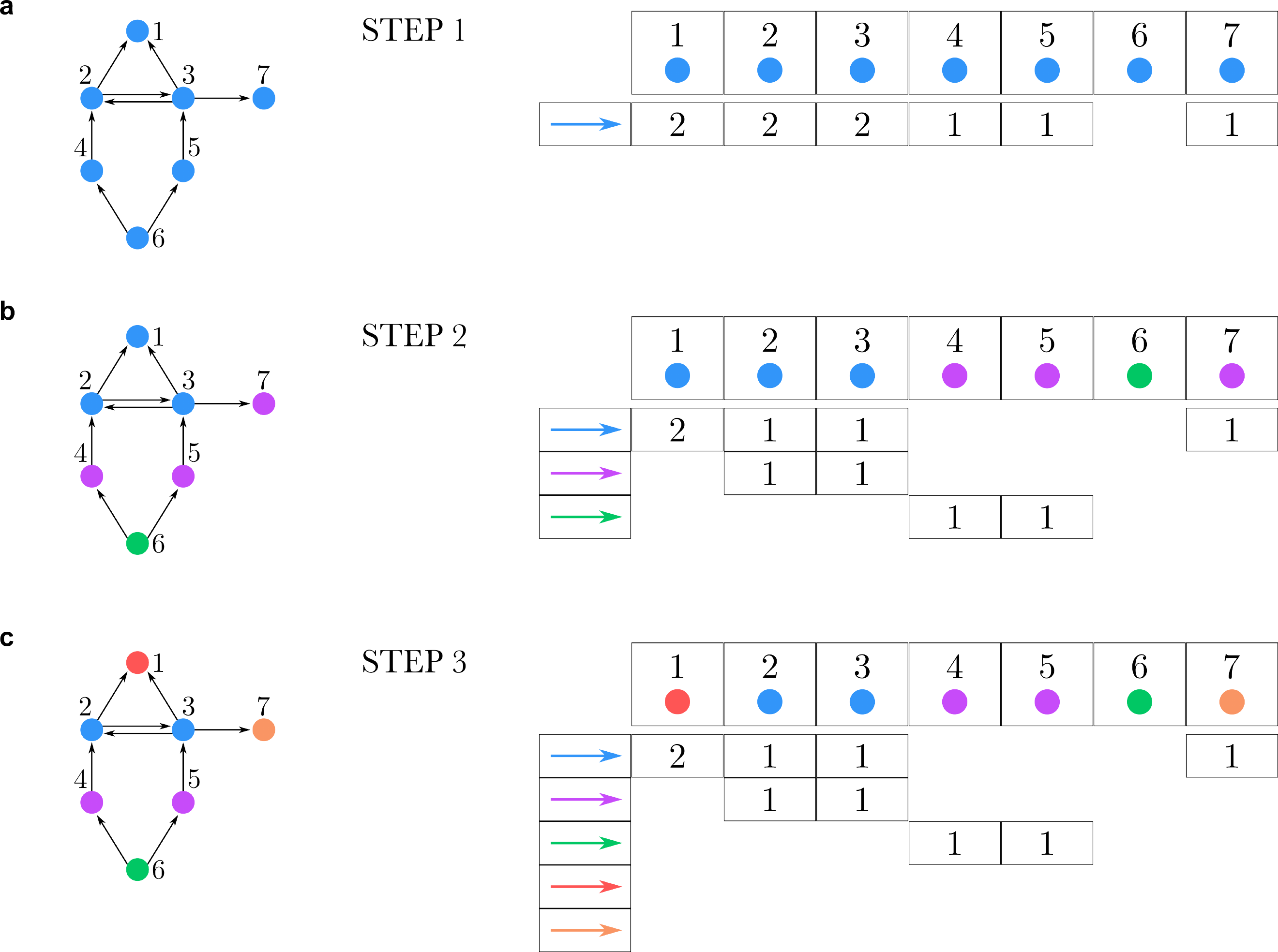} 
\centering
\caption{{\bf Algorithm to find the fibers of a network through a
   minimal balanced coloring.} The goal of the algorithm is to find a minimal 
    balanced coloring of the network, so that two nodes have the same
  color only if they are connected to the same number of identically
  colored nodes via inbound links. The colors represent the fibers in
  the network.}
\label{fig:algorithm}
\end{figure}

The algorithm constructs a coloring of the nodes that is balanced. A
coloring is balanced if two identically colored nodes are connected to
identically colored nodes via their inbound links.
Each balanced colored cluster is a fiber in the network. The fibers
also corresponds to the orbits in a network when the symmetries are
automorphisms rather than isomorphisms in the input trees.  The flow
of the algorithm is exemplified with the example network of
Fig.~\ref{fig:algorithm}.

\begin{itemize}
\item {\bf Step 1} - We start by assigning the same color to all nodes.  In
  Fig.~\ref{fig:algorithm}a all nodes are initially colored in blue.
  In addition, we assign to each link the same color of the node from
  where it emanates.  To update the coloring (or, equivalently, to
  generate a new partition) of nodes, we construct the table shown in
  the right panel of Fig.~\ref{fig:algorithm}a, as explained next.  In
  the top row of this table we put the network nodes colored with
  their current color.  In the leftmost column we put each type of
  colored link. In this initial stage of the algorithm we only have a
  blue link for all the nodes.  Then, we fill the entries of the table
  with the number of colored links of this blue type that are received
  by the corresponding node.  For example, node 1 receives two 2 blue
  links as well as nodes 2 and 3. Nodes 4, 5 and 7 receive one blue
  link each, and node 6 nothing. The structure of this table
  determines the new coloring as explained in the next step.
\item {\bf Step 2} - Using the table in~Fig.~\ref{fig:algorithm}a we update
  the coloring of nodes as follows. We assign the same color to all
  nodes that receive the same number of colored links of each
  type. Specifically, nodes 1, 2 and 3 receive two blue links, so we
  assign them the same (blue) color. Analogously, nodes 4, 5 and 7
  receive one blue link, so we assign them the same color, but
  different from blue. We assign them a purple color. Similarly, we
  assign another color to node 6 (green).  We then obtain the colored
  network in the left of Fig. \ref{fig:algorithm}b. Applying the
  counting of receiving coloring links to this network, we obtain the
  new coloring table shown in~Fig.~\ref{fig:algorithm}b, where each
  link has the color of the node from where it emanates.  Thus, we
  update the table to generate the new coloring, as shown in the right
  panel of~Fig.~\ref{fig:algorithm}b.
%
%
%
\item {\bf Step 3} - Using the same criterion as in Step 2, we update the
  coloring of nodes, comprising now five different colors, and then we
  generate the new table, as shown in~Fig.~\ref{fig:algorithm}c.  At
  this point the algorithm stops, because we do not need to introduce
  more colors, since each color is balanced. Each color corresponds to
  a fiber, and each node in each colored fiber receives the same
  colors from other fibers or from nodes in the same fiber. Therefore,
  the coloring shown in the network of Fig.~\ref{fig:algorithm}c is
  the minimal balanced coloring of the network, and the colors
  indicate the fibers in the network.
\end{itemize}

As far as only minimal fibrations are considered, the algorithm will
return always the same fibers containing the same nodes, for any
initial condition and realization.
Below we provide the pseudo-code to clarify the
  algorithm. More detailed instructions and methodology for obtaining
  fiber building blocks will be given in a follow-up paper.  We start
  by assigning all nodes to the same fiber and then continue to refine
  the partition basing on the input set of the node until no further
  refinement can be obtained.

\begin{algorithm}
  \caption{Finding fibers following Kamei \& Cock Ref. \cite{kamei}}
  \label{pseudoKamei}
  \begin{flushleft}
    \hspace*{\algorithmicindent} \textbf{Input:} Graph $G = \{N_G, E_G\}$, where $N_G$ are vertices and $E_G$ are edges of the analyzed network \\
    \hspace*{\algorithmicindent} $\hskip 3em$ $\mid N_G \mid$ - number of vertices, $N_G = \{v_1 \dots v_{\mid N_G \mid}\}$ \\
    \hspace*{\algorithmicindent} \textbf{Output:} $C = \{c_i\}$, where $c_i$ - color of node $i$ and $i = 1 \dots \mid V \mid$ \\
    \hspace*{\algorithmicindent} \textbf{Notation:} $I_i = \{I_i^1 \dots I_i^N\}$, where N = current number of colors
  \end{flushleft}
  
  \begin{algorithmic}[1]
    \State $N_0 = 1$
    \For{$i = 1 \dots \mid N_G \mid$}
      \State $c_i = 1$
    \EndFor
    \State $j = 0$
    \Repeat
      \For{$i = 1 \dots \mid N_G \mid$, $k = 1...N_j$}
        \State $I_i^k = $ number of nodes of color $k$ in the input set of $v_i$
      \EndFor
      \State $H$ = set of all unique $\{I_i\}$
      \State // assign each unique vector a color and color the graph accordingly
      \For{$i = 1 \dots \mid N_G \mid$}
        \State $c_i = $ index of $I_i$ in $H$, e.g. if two nodes have the same $I_i$ and $I_j$ $\rightarrow$ $c_i = c_j$
      \EndFor
      \State $j = j + 1$
      \State $N_j = \, \mid H \mid$
    \Until{$N_j \neq N_{j - 1}$}
    \State \Return{$\{c_i\}$}
  \end{algorithmic}
\end{algorithm}

\clearpage

\section{Strongly connected component}
\label{gcc}

\begin{figure*}
  \includegraphics[width=.7\textwidth]{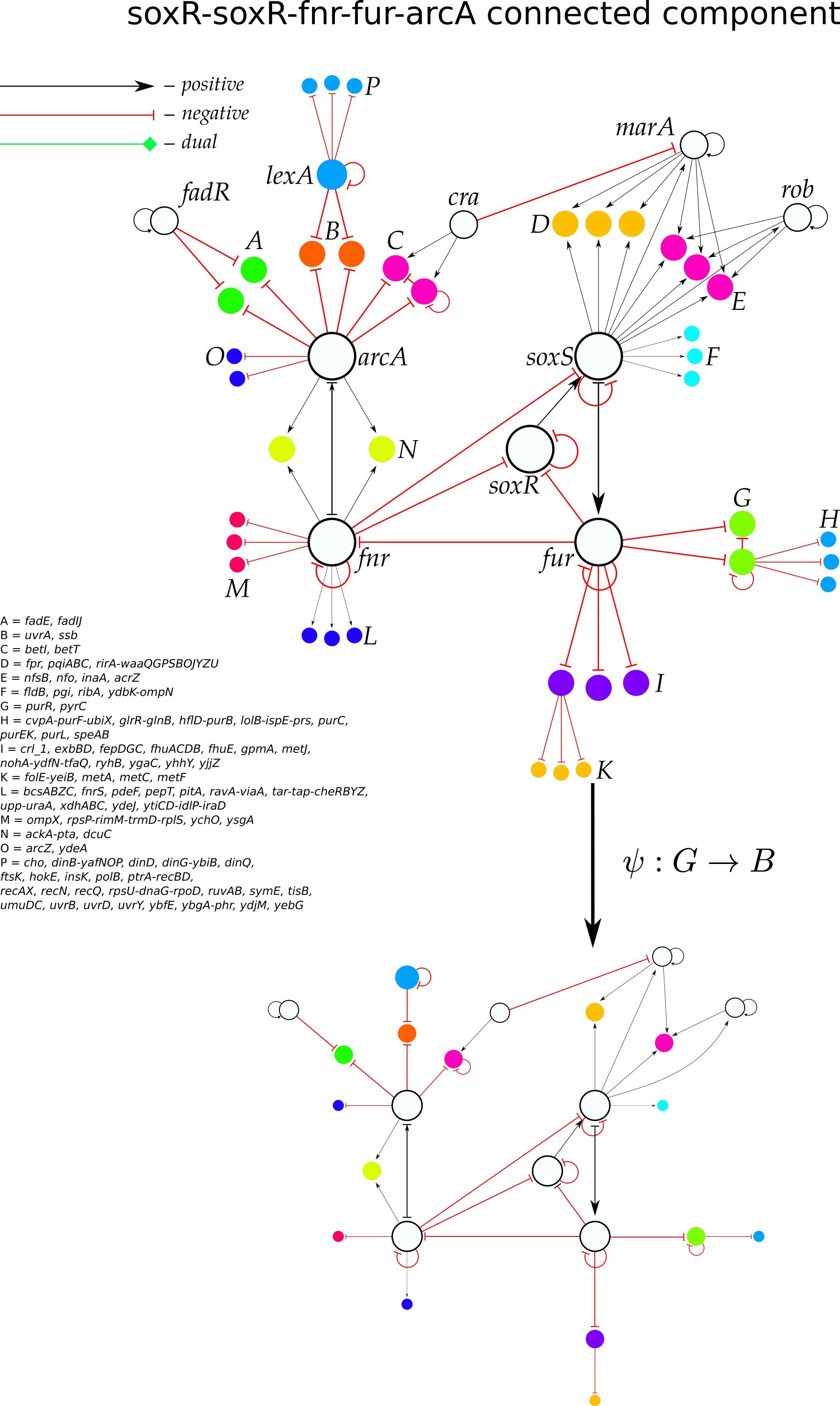} \centering
  \caption{A five-gene connected component of {\it soxR, soxS, fnr,
      fur, } and {\it arcA} with its regulated fibers.}
\label{soxr}
\end{figure*}

In a directed network, the strongly connected component is composed of
nodes that are reachable from every other node in the component.  That
is, there is a directed path from every node to any other node in the
strongly connected component. A weakly connected component is obtained
when we ignore the directionality of the links. Strongly connected
components are relevant to genetic fibers since they contain loops
that control the state of the genes.  We find four types of strongly
connected components. Single-gene components composed of autoregulator
loops like {\it cpxR} and {\it fadR} in Figs. \ref{cpxr}a and
\ref{cpxr}e. The other type of components are those in
Fig. \ref{components}a and Fig. \ref{components}b and also a five-gene
connected component shown in SI Fig. \ref{soxr}. We note that most of
the fibers regulated by these components do not belong to the
connected component. This is because they receive information but do
not send information back to the connected component. These fibers are
characterized by integer fiber numbers.  When the fiber receives and
sends back information, that is, when the fiber belongs to the
strongly connected component, then it becomes a Fibonacci fiber.  The
largest strongly connected component in the {\it E. coli } network
controls the pH system shown in Fig. \ref{components}b.

\section{Statistics of fibers in the TRN of {\it E. coli}}
\label{stats}

\subsection{Fibers statistics in {\it E. coli}}

SI Table \ref{statistics} shows the counts in the {\it E. coli}
network of each building block. For instance the most abundant
building blocks are the following:\\

\noindent \textbf{$\rvert n = 0, \ell = 1 \rangle$}: 45  \\
 \textbf{$\rvert n = 1, \ell = 0 \rangle$}:  13 \\
   \textbf{$\rvert n = 0, \ell = 2 \rangle$}:  13 \\
   \textbf{$\rvert n = 1, \ell = 1 \rangle$}: 8 \\

The list is completed with the fractal building blocks of Fibonacci
sequences which are less numerous but more complex in their structure:
   
\noindent \textbf{$\rvert \varphi_2 = 1.6180.., \ell = 2 \rangle$}: 1
\\ \textbf{$\rvert \varphi_3 = 1.4655.., \ell = 1 \rangle$}: 1
\\ \textbf{$\rvert \varphi_4 = 1.3802..., \ell = 1 \rangle$}: 1 \\

\begin{table*}[ht]
\centering
\begin{tabular}{| c | c |}
  \hline Structure type & Amount in E-coli \\
  \hline
  \textbf{$\rvert n = 0, l = 1 \rangle$} & 45 \\
  \textbf{$\rvert n = 0, l = 2 \rangle$} & 13 \\
  \textbf{$\rvert n = 0, l = 3 \rangle$} & 3 \\
  \textbf{$\rvert n = 1, l = 0 \rangle$} & 13 \\
  \textbf{$\rvert n = 1, l = 1 \rangle$} & 8 \\
  \textbf{$\rvert n = 1, l = 2 \rangle$} & 3 \\
  \textbf{$\rvert n = 2, l = 0 \rangle$} & 1 \\
  \textbf{$\rvert n = 2, l = 1 \rangle$} & 1 \\
  \textbf{$\rvert \varphi_d = 1.3802.., l = 1 \rangle$} & 1 \\
  \textbf{$\rvert \varphi_d = 1.4655.., l = 1 \rangle$} & 1 \\
  \textbf{$\rvert \varphi_d = 1.6180.., l = 2 \rangle$} & 1 \\
  Composite Fiber & 1 \\
  \hline
  \textbf{Total number of building blocks}  & 91 \\
  \hline
\end{tabular}
\caption{Building block statistics. We show the count of every
  building block defined by the fiber numbers.}
\label{statistics}
\end{table*}

\subsection{Full list of fibers in {\it E. coli}}

SI Table \ref{all_building_blocks} shows the complete list of the 91
fibers building blocks found in the genetic network of {\it
  E. coli}. We list the genes in the fiber plus their external
regulators. If a gene or operon is not in this list, for instance {\it
  lacZYA}, it means that the gene or operon is not in a fiber.
Supplementary File 1 shows the plot of the circuit of every fiber and
the fiber building block.

The first column in SI Table \ref{all_building_blocks} is the ID of
the fiber. This ID refers to the plot of the fiber building block in
Supplementary File 1. The second column lists the genes in the fiber,
the third column lists the external regulators. The last column
specifies the fiber number associated with each fiber as $\rvert n,
\ell \rangle$
or
$\rvert \varphi_d, \ell \rangle$.

\section{Datasets of biological and non-biological networks}
\label{datasets}

To investigate the applicability of fibrations in a broader context,
we performed an extensive analysis of different complex networks from
diverse domains in systems science.

Full details of each network analyzed can be accessed at
\url{https://docs.google.com/spreadsheets/d/1-RG5vR_EGNPqQcnJU8q3ky1OpWi3OjTh5Uo-Xa0PjOc}. The
codes to reproduce this analysis are at \url{github.com/makselab} and
the full datasets appear at \url{kcorelab.org}. See also tables below
with information about the networks.

We first show the symmetry fibrations in biological
  networks and species.  See Section \ref{bio}. We characterize
biological networks spanning from:

\begin{itemize}
\item {\bf Biological networks: transcriptional regulatory networks,
  metabolic networks, cellular processes networks and pathways,
  disease networks, neural networks.}
\end{itemize}

We study the following species:

\begin{itemize}
\item {\bf Species: A. thaliana, E. coli, B. subtilis, S. enterica
  (salmonella), M. tuberculosis, D. melanogaster, S. cerevisiae
  (yeast), M. musculus (mouse), and H. sapiens (human).}

\end{itemize}

We then study non-biological networks in Section
  \ref{bio}:
    
\begin{itemize}

\item {\bf Social Networks: online social networks, Facebook, Twitter,
  Wikipedia, Youtube, email networks, communication networks, citation
  networks, collaboration networks, bloggers }

\item {\bf Internet: routers, autonomous systems, web graphs,
  hyperlinks, peer-to-peer}

\item {\bf Infrastructure Networks: power grid, airport, roads,
  flights}
  
\item {\bf Economic Networks }

\item {\bf Software Networks: Linux, jdk}

\item {\bf Ecosystems }

\end{itemize}

  \begin{table}[ht]
  \centering
  \begin{tabular}{|l|c|c|c|}
    \hline
    Network Domain & Total No. of nodes & Total No. of edges & No. of networks \\ 
    \hline
 Biological & 287390 & 4211856 & 289 \\ 
 Economic & 1752 & 108639 &   5 \\ 
 Ecosystems & 1879 & 5378 &  14 \\ 
 Infrastructure & 24511 & 82534 &  16 \\ 
 Internet & 244634 & 835565 &  27 \\ 
 Social & 104909 & 1261009 &  15 \\ 
 Software & 43391 & 503645 &   3 \\ 
     \hline
  \end{tabular}
\caption{Features of the networks across domains. We report the total
  numbers for each domain summed over all the networks in the domain.}
  \end{table}

\begin{table}[ht]
\centering
\begin{tabular}{|l|c|c|c|}
  \hline
 Species & Total No. of nodes & Total No. of edges & No. networks \\ 
  \hline
  Yeast & 55932 & 1392926 &  11 \\ 
  Arabidopsis Thaliana & 790 & 1431 &   1 \\ 
  Bacillus subtilis & 5602 & 11417 &   3 \\ 
  Drosophila & 39549 & 321734 &   5 \\ 
  Escherichia coli & 879 & 1835 &   1 \\ 
  Human & 72587 & 1198712 & 248 \\ 
  Micobacterium Tuberculosis & 1624 & 3212 &   1 \\ 
  Mouse & 64709 & 987424 &   7 \\ 
  Salmonella & 8293 & 15589 &   6 \\ 
   \hline
\end{tabular}
\caption{Number of networks per species.}
\end{table}

\begin{table}[ht]
\centering
\resizebox{\textwidth}{!}{\begin{tabular}{| c | c | c | c | c | c | c | c | c | c | c | c | c |}
  \hline
    & Arabidopsis & Bacillus & Caenorhabditis & Cat & Drosophila & Escherichia & Human & Micobacterium & Mouse & Rat & Salmonella & Yeast \\
    & Thaliana & subtilis & elegans & & & coli & & Tuberculosis & & & & \\ 
  \hline
    TF & 1 & 2 & 2 & 0 & 4 & 1 & 4 & 1 & 4 & 0 & 2 & 11 \\
    Neuron & 0 & 0 & 0 & 1 & 1 & 0 & 0 & 0 & 3 & 3 & 0 & 0 \\ 
    Metabolic & 0 & 0 & 0 & 0 & 0 & 0 & 48 & 0 & 0 & 0 & 2 & 0 \\ 
    Disease & 0 & 0 & 0 & 0 & 0 & 0 & 66 & 0 & 0 & 0 & 0 & 0 \\ 
    Kinase & 0 & 0 & 0 & 0 & 0 & 0 & 2 & 0 & 0 & 0 & 0 & 0 \\ 
    Pathway & 0 & 0 & 0 & 0 & 0 & 0 & 127 & 0 & 0 & 0 & 0 & 0 \\ 
    Protein & 0 & 1 & 0 & 0 & 0 & 0 & 1 & 0 & 0 & 0 & 2 & 0 \\ 
   \hline
\end{tabular}}
\caption{Table with the count of networks per type of biological
  network and species. These networks are used to calculate the
  distributions of fiber across species and biological types in
  Figs. \ref{reduction}a, b, and c. For each type of biological
  network in Fig. \ref{reduction}a, b, we calculate the count over the
  total number of networks as indicates at the end of each row for
  each biological type. The same occurs with the number of networks at
  the end of each column for each species. Figure \ref{reduction}c
  shows the counts over all the network shown in the last row/column.}
\label{networks}
\end{table}

\begin{table}[ht]
\centering
\begin{tabular}{|l|c|c|c|}
  \hline
 Network Subdomain & Total No. of nodes & Total No. of edges & No. of networks \\ 
  \hline
  Autonomous systems graphs & 141842 & 481415 &  14 \\ 
  Bitcoin & 9664 & 59777 &   2 \\ 
  Collaboration networks & 50260 & 504897 &   4 \\ 
  Disease & 4309 & 15254 &  66 \\ 
  Facebook & 4039 & 88234 &   1 \\ 
  Youtube subscriptions & 13723 & 76765 &   1 \\ 
  Internet peer-to-peer networks & 31978 & 110154 &   4 \\ 
  Jazz & 198 & 5484 &   1 \\ 
  Linux & 30837 & 213954 &   1 \\ 
  Metabolic & 4273 & 33829 &  50 \\ 
  Networks with ground-truth communities & 1005 & 25571 &   1 \\ 
  Neural networks & 3694 & 129812 &   8 \\ 
  Cellular processes and Pathways & 9825 & 54712 & 127 \\ 
  Plant-Pollinator & 1631 & 2719 &  11 \\ 
  Plant-Seed-Disperser &  65 & 165 &   2 \\ 
  Power grid & 4941 & 6594 &   1 \\ 
  Sentiment &  99 & 278 &   2 \\ 
  Transcriptional regulatory & 260258 & 3908769 &  32 \\ 
   \hline
\end{tabular}
\caption{Subtypes of networks belonging to the different domains.}
\end{table}

\clearpage

\begin{longtable}{| c | p{10cm} | p{2cm} | H p{3.5cm} |}
  \hline
Id & Fiber & Regulators & Block Name & Fiber Number \\ 
  \hline
    1 & aaeR, ampDE, azuC, comR, cyaA, narQ, sohB, speC, spf, trxA, yaeP-rof, yaeQ-arfB-nlpE, yjeF-tsaE-amiB-mutL-miaA-hfq-hflXKC & crp & 13-USF & $\rvert n = 0, l = 1 \rangle$ \\ 
    2 & aaeXAB, agp, cpdB, cstA, glgS, glpR, grpE, hofMNOP, ivbL-ilvBN-uhpABC, lacI, mcaS, mhpR, nadC, ompA, ppdD-hofBC, preTA, raiA, rmf, rpsF-priB-rpsR-rplI, sfsA-dksA-gluQ, sxy, ubiG, ychH, yeiP, yeiW, yfiP-patZ, yibN-grxC-secB-gpsA, ykgR & crp & 28-USF & $\rvert n = 0, l = 1 \rangle$ \\ 
    3 & accA, accD, fabI, fadR, yceD-rpmF-plsX-fabHDG-acpP-fabF &  & 4-ASF & $\rvert n = 1, l = 0 \rangle$ \\ 
    4 & accB, iclR & fadR & 1-2-FFF & $\rvert n = 1, l = 1 \rangle$ \\ 
    5 & ackA-pta, dcuC & arcA, fnr & 2-2-FAN & $\rvert n = 0, l = 2 \rangle$ \\ 
    6 & acrZ, inaA, nfo, nfsB & marA, rob, soxS & 3-4-FAN & $\rvert n = 0, l = 3 \rangle$ \\ 
    7 & add, dsbG, gor, grxA, hemH, oxyS, trxC & crp, oxyR, rbsR & Composite Fiber & $\rvert n = 0, l = 1 \rangle \oplus \rvert n = 1, l = 1 \rangle$ \\ 
    8 & adeD, adiY, chiA, gspAB, hchA, hdfR, mdtJI, rcsB, yjjP & hns & 2-FF & $\rvert \varphi_d = 1.4655.., l = 1 \rangle$ \\ 
    9 & agaR, agaS-kbaY-agaBCDI &  & 2-CF & $\rvert n = 1, l = 0 \rangle$ \\ 
   10 & alaA-yfbR, avtA, leuE, livJ, livKHMGF, lysU, sdaA & lrp & 7-USF & $\rvert n = 0, l = 1 \rangle$ \\ 
   11 & alaE, kbl-tdh, yojI & lrp & 3-USF & $\rvert n = 0, l = 1 \rangle$ \\ 
   12 & alaWX, argU, argW, argX-hisR-leuT-proM, aspV, flxA, glyU, leuQPV, leuX, lptD-surA-pdxA-rsmA-apaGH, lysT-valT-lysW, metT-leuW-glnUW-metU-glnVX, pheU, pheV, proK, proL, queA, serT, serX, thrU-tyrU-glyT-thrT-tufB, thrW, trmA, tyrTV-tpr, valUXY-lysV & fis & 24-USF & $\rvert n = 0, l = 1 \rangle$ \\ 
   13 & aldB, hupB & crp, fis & 2-2-FAN & $\rvert n = 0, l = 2 \rangle$ \\ 
   14 & allA, allS, gcl-hyi-glxR-ybbW-allB-ybbY-glxK & allR & 3-USF & $\rvert n = 0, l = 1 \rangle$ \\ 
   15 & alsR, rpiB &  & 2-CF & $\rvert n = 1, l = 0 \rangle$ \\ 
   16 & amiA-hemF, cmk-rpsA-ihfB, uspB & IHF & 3-USF & $\rvert n = 0, l = 1 \rangle$ \\ 
   17 & amn, mipA, phnCDE\_1E\_2FGHIJKLMNOP, phoA-psiF, phoB, phoE, phoH, ydfH, yegH, yhjC, ytfK &  & 10-ASF & $\rvert n = 1, l = 0 \rangle$ \\ 
   18 & ampC, dacC & bolA & 2-USF & $\rvert n = 0, l = 1 \rangle$ \\ 
   19 & araE-ygeA, araFGH & araC, crp & 2-2-FAN & $\rvert n = 0, l = 2 \rangle$ \\ 
   20 & arcZ, ydeA & arcA & 2-USF & $\rvert n = 0, l = 1 \rangle$ \\ 
   21 & argA, argCBH, argE, argF, argI, argR, artJ, artPIQM, lysO &  & 8-ASF & $\rvert n = 1, l = 0 \rangle$ \\ 
   22 & argO, lysP & argP, lrp & 2-2-FAN & $\rvert n = 0, l = 2 \rangle$ \\ 
   23 & aroF-tyrA, tyrB & tyrR & 2-USF & $\rvert n = 0, l = 1 \rangle$ \\ 
   24 & aroH, trpLEDCBA, trpR &  & 2-ASF & $\rvert n = 1, l = 0 \rangle$ \\ 
   25 & asnB, clpPX-lon, glsA-ybaT, uspE & gadX & 4-USF & $\rvert n = 0, l = 1 \rangle$ \\ 
   26 & aspA-dcuA, dcuR & crp, fnr, narL & 3-2-FAN & $\rvert n = 0, l = 3 \rangle$ \\ 
   27 & bacA, cpxPQ, cpxR, ftnB, ldtC, ldtD, ppiD, sbmA-yaiW, slt, srkA-dsbA, xerD-dsbC-recJ-prfB-lysS, yccA, yebE, yidQ, yqaE-kbp, yqjA-mzrA &  & 15-ASF & $\rvert n = 1, l = 0 \rangle$ \\ 
   28 & baeR, spy & cpxR & 1-2-FFF & $\rvert n = 1, l = 1 \rangle$ \\ 
   29 & bcsABZC, fnrS, pdeF, pepT, pitA, ravA-viaA, tar-tap-cheRBYZ, upp-uraA, xdhABC, ydeJ, ytiCD-idlP-iraD & fnr & 11-USF & $\rvert n = 0, l = 1 \rangle$ \\ 
   30 & bdcA, dkgB, grxD, mepH, mhpT, pgpC-tadA, rfe-wzzE-wecBC-rffGHC-wecE-wzxE-rffT-wzyE-rffM, rybB, tehAB, tsgA, ydbD, yeaE & nsrR & 12-USF & $\rvert n = 0, l = 1 \rangle$ \\ 
   31 & betI, betT & arcA, cra & 2-2-FFF & $\rvert n = 1, l = 2 \rangle$ \\ 
   32 & bioA, bioBFCD & birA & 2-USF & $\rvert n = 0, l = 1 \rangle$ \\ 
   33 & bluF, ydeI & rcdA & 2-USF & $\rvert n = 0, l = 1 \rangle$ \\ 
   34 & borD, envY-ompT, mgrB, mgrR, mgtLA, mgtS, pagP, rstA, ybjG & phoP & 9-USF & $\rvert n = 0, l = 1 \rangle$ \\ 
   35 & cbpAM, gltX, gyrB, msrA & fis & 4-USF & $\rvert n = 0, l = 1 \rangle$ \\ 
   36 & cdaR, garD, gudPXD &  & 2-ASF & $\rvert n = 1, l = 0 \rangle$ \\ 
   37 & cho, dinB-yafNOP, dinD, ding-ybib, dinQ, ftsK, hokE, insK, lexA, polB, ptrA-recBD, recAX, recN, recQ, rpsU-dnaG-rpoD, ruvAB, symE, tisB, umuDC, uvrB, uvrD, uvrY, ybfE, ybgA-phr, ydjM, yebG &  & 25-ASF & $\rvert n = 1, l = 0 \rangle$ \\ 
   38 & cirA, entCEBAH, fepA-entD, fiu & crp, fur & 2-4-FAN & $\rvert n = 0, l = 2 \rangle$ \\ 
   39 & copA, cueO & cueR & 2-USF & $\rvert n = 0, l = 1 \rangle$ \\ 
   40 & cra, pitB, sbcDC & phoB & 3-USF & $\rvert n = 0, l = 1 \rangle$ \\ 
   41 & crl\_1, exbBD, fepDGC, fhuACDB, fhuE, gpmA, metJ, nohA-ydfN-tfaQ, ryhB, ygaC, yhhY, yjjZ & fur & 12-USF & $\rvert n = 0, l = 1 \rangle$ \\ 
   42 & cusCFBA, cusR, yedX & hprR, phoB & 2-3-FFF & $\rvert n = 1, l = 2 \rangle$ \\ 
   43 & cvpA-purF-ubiX, glrR-glnB, hflD-purB, lolB-ispE-prs, purC, purEK, purL, speAB & purR & 8-USF & $\rvert n = 0, l = 1 \rangle$ \\ 
   44 & cysDNC, cysK, tcyP, yciW, ygeH, yoaC & cysB & 6-USF & $\rvert n = 0, l = 1 \rangle$ \\ 
   45 & cytR, nagC, nagE, ycdZ & crp & 1-4-FFF & $\rvert n = 1, l = 1 \rangle$ \\ 
   46 & dapB, lysC & argP & 2-USF & $\rvert n = 0, l = 1 \rangle$ \\ 
   47 & ddpXABCDF, patA, potFGHI, yeaGH, yhdWXYZ & ntrC & 5-USF & $\rvert n = 0, l = 1 \rangle$ \\ 
   48 & decR, mlaFEDCB, yncE & marA & 3-USF & $\rvert n = 0, l = 1 \rangle$ \\ 
   49 & dgcC, iraP, nlpA, wrbA-yccJ, yccT & csgD & 5-USF & $\rvert n = 0, l = 1 \rangle$ \\ 
   50 & dicB-ydfDE-insD-7-intQ, dicC-ydfXW & dicA & 2-USF & $\rvert n = 0, l = 1 \rangle$ \\ 
   51 & dsdC, norR & nsrR & 1-2-FFF & $\rvert n = 1, l = 1 \rangle$ \\ 
   52 & dtpA, omrA, omrB & ompR & 3-USF & $\rvert n = 0, l = 1 \rangle$ \\ 
   53 & ecpA, ecpR & matA & 2-USF & $\rvert n = 0, l = 1 \rangle$ \\ 
   54 & efeU\_1U\_2, motAB-cheAW, psd-mscM, tsr, ung & cpxR & 5-USF & $\rvert n = 0, l = 1 \rangle$ \\ 
   55 & epd-pgk-fbaA, gapA-yeaD, mpl & cra, crp & 2-3-FAN & $\rvert n = 0, l = 2 \rangle$ \\ 
   56 & erpA, iscR, rnlAB &  & 2-ASF & $\rvert n = 1, l = 0 \rangle$ \\ 
   57 & evgA, nhaR & hns & 3-FF & $\rvert \varphi_d = 1.3802.., l = 1 \rangle$ \\ 
   58 & fabA, fabB & fabR, fadR & 2-2-FAN & $\rvert n = 0, l = 2 \rangle$ \\ 
   59 & fadE, fadIJ & arcA, fadR & 2-2-FAN & $\rvert n = 0, l = 2 \rangle$ \\ 
   60 & fbaB, fruBKA, glk, gpmM-envC-yibQ, pfkA, ppc, pykF, pyrG-eno, tpiA & cra & 9-USF & $\rvert n = 0, l = 1 \rangle$ \\ 
   61 & fldB, pgi, ribA, ydbK-ompN & soxS & 4-USF & $\rvert n = 0, l = 1 \rangle$ \\ 
   62 & folE-yeiB, metA, metC, metF & metJ & 4-USF & $\rvert n = 0, l = 1 \rangle$ \\ 
   63 & fpr, pqiABC, rirA-waaQGPSBOJYZU & marA, soxS & 2-3-FAN & $\rvert n = 0, l = 2 \rangle$ \\ 
   64 & fucAO, fucR, zraR & crp & 1-3-FFF & $\rvert n = 1, l = 1 \rangle$ \\ 
   65 & gfcA, ybhL, yfiR-dgcN-yfiB, ymiA-yciX & yjjQ & 4-USF & $\rvert n = 0, l = 1 \rangle$ \\ 
   66 & hupA, trg & crp, fis & 2-2-FAN & $\rvert n = 0, l = 2 \rangle$ \\ 
   67 & ibaG-murA, rplU-rpmA-yhbE-obgE & mlrA & 2-USF & $\rvert n = 0, l = 1 \rangle$ \\ 
   68 & ibpAB, yadV-htrE & IHF & 2-USF & $\rvert n = 0, l = 1 \rangle$ \\ 
   69 & idnK, idnR & crp, gntR & 2-2-FFF & $\rvert n = 1, l = 2 \rangle$ \\ 
   70 & isrC-flu, pth-ychF & oxyR & 2-USF & $\rvert n = 0, l = 1 \rangle$ \\ 
   71 & lgoR, uxuR & crp, exuR & 2-FF & $\rvert \varphi_d = 1.6180.., l = 2 \rangle$ \\ 
   72 & lolA-rarA, osmB & rcsB & 2-USF & $\rvert n = 0, l = 1 \rangle$ \\ 
   73 & lsrACDBFG-tam, lsrR, oxyR, rbsR & crp & 1-4-FFF & $\rvert n = 1, l = 1 \rangle$ \\ 
   74 & malI, mlc & crp & 1-2-FFF & $\rvert n = 1, l = 1 \rangle$ \\ 
   75 & manA, yhfA & crp & 1-2-FAN & $\rvert n = 0, l = 1 \rangle$ \\ 
   76 & mngAB, mngR &  & 2-CF & $\rvert n = 1, l = 0 \rangle$ \\ 
   77 & nadA-pnuC, nadB & nadR & 2-USF & $\rvert n = 0, l = 1 \rangle$ \\ 
   78 & nimR, nimT &  & 2-CF & $\rvert n = 1, l = 0 \rangle$ \\ 
   79 & ompX, rpsP-rimM-trmD-rplS, ychO, ysgA & fnr & 4-USF & $\rvert n = 0, l = 1 \rangle$ \\ 
   80 & pepD, yhbTS & csgD & 2-USF & $\rvert n = 0, l = 1 \rangle$ \\ 
   81 & phoP, slyB &  & 2-CF & $\rvert n = 2, l = 0 \rangle$ \\ 
   82 & pspABCDE, pspG & IHF, pspF & 2-2-FAN & $\rvert n = 0, l = 2 \rangle$ \\ 
   83 & purR, pyrC & fur & 1-2-FFF & $\rvert n = 1, l = 1 \rangle$ \\ 
   84 & rhaR, rhaS & crp & 1-2-BTF & $\rvert n = 2, l = 1 \rangle$ \\ 
   85 & rrsA-ileT-alaT-rrlA-rrfA, rrsE-gltV-rrlE-rrfE & fis, lrp & 2-2-FAN & $\rvert n = 0, l = 2 \rangle$ \\ 
   86 & rrsB-gltT-rrlB-rrfB, rrsC-gltU-rrlC-rrfC, rrsD-ileU-alaU-rrlD-rrfD-thrV-rrfF, rrsG-gltW-rrlG-rrfG, rrsH-ileV-alaV-rrlH-rrfH & fis, hns, lrp & 3-5-FAN & $\rvert n = 0, l = 3 \rangle$ \\ 
   87 & ssb, uvrA & arcA, lexA & 2-2-FAN & $\rvert n = 0, l = 2 \rangle$ \\ 
   88 & ttdABT, ttdR &  & 2-CF & $\rvert n = 1, l = 0 \rangle$ \\ 
   89 & ycjG, ycjY-ymjDC-mpaA & pgrR & 2-USF & $\rvert n = 0, l = 1 \rangle$ \\ 
   90 & yegRZ, yfdX-frc-oxc-yfdVE & evgA & 2-USF & $\rvert n = 0, l = 1 \rangle$ \\ 
   91 & ykgMO, znuA, znuCB & zur & 3-USF & $\rvert n = 0, l = 1 \rangle$ \\ 
   \hline
  \caption{List of fiber building blocks with ID, 
    genes in the fiber, external regulators of the fiber
    and  fiber numbers. We provide 
    Supplementary File 1 which plots every building block
    using the same IDs. }
  \label{all_building_blocks}
\end{longtable}

\end{document}